\documentclass[longbibliography,aps,reprint,groupedaddress,superscriptaddress]{revtex4-1}

\setcitestyle{super,open={},close={}}
\usepackage{lineno}
\usepackage{graphicx}
\usepackage{bm}
\usepackage{amsmath,amssymb}
\usepackage{hyperref}
\usepackage{lipsum}
\usepackage{color}
\definecolor{colorref}{rgb}{0.0, 0.408, 0.647}
\definecolor{grey}{rgb}{0.95, 0.95, 0.95}
\hypersetup{
	colorlinks = true,
	allcolors = colorref
}
\newcommand{\figtitle}[1]{\textbf{#1}}
\newcommand{\subfiglabel}[1]{\textbf{#1}}
\newcommand{\rsubfiglabel}[1]{(\textbf{#1})}

\newcommand{\rfig}[1]{Fig.~\textcolor{colorref}{\ref{#1}}}

\newcommand{\rsubfig}[2]{Fig.~\textcolor{colorref}{\ref{#1}#2}}
\newcommand{\rsubfigs}[3]{Fig.~\textcolor{colorref}{\ref{#1}#2}-\textcolor{colorref}{#3}}
\newcommand{\rsubfigsA}[3]{Fig.~\textcolor{colorref}{\ref{#1}#2, #3}}

\newcommand{\SeeSupply}[1]{Supplemental Information}

\definecolor{reviseColor}{rgb}{1, 0.0, 0.0}

\usepackage{pdfpages} 
\usepackage{pgffor} 
\makeatletter
\AtBeginDocument{\let\LS@rot\@undefined}
\makeatother
\def\supplementfilename{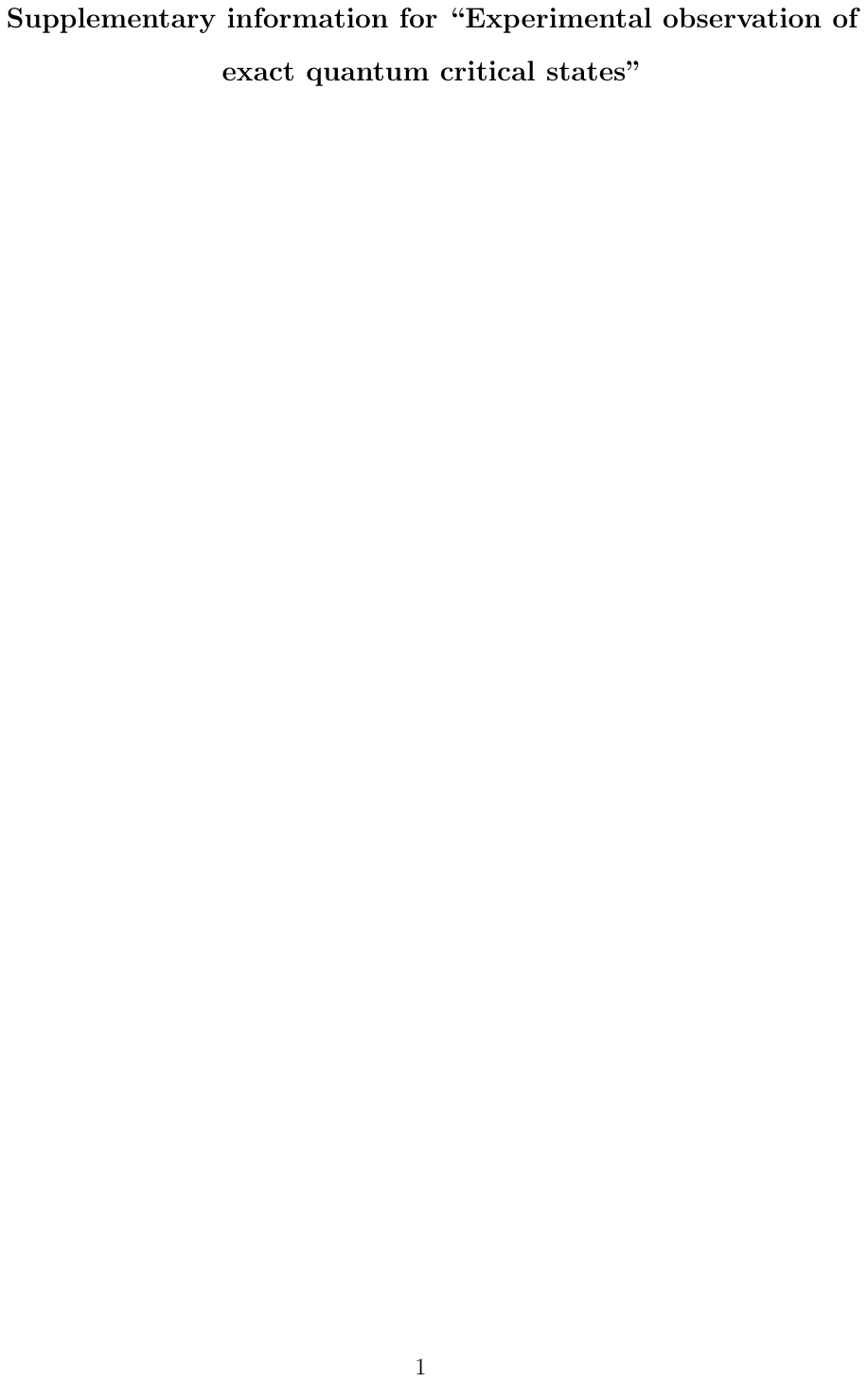}
\pdfximage{\supplementfilename}
\def\numbersupplementpages{\the\pdflastximagepages}

\nolinenumbers

\newcommand{\SIQSE}{\affiliation{1}{Shenzhen Institute for Quantum Science and Engineering, Southern University of Science and Technology, Shenzhen 518055, China}}
\newcommand{\PKU}{\affiliation{2}{International Center for Quantum Materials and School of Physics, Peking University, Beijing 100871, China}}
\newcommand{\IQA}{\affiliation{3}{International Quantum Academy, Shenzhen 518048, China}}
\newcommand{\GDKL}{\affiliation{4}{Guangdong Provincial Key Laboratory of Quantum Science and Engineering, Southern University of Science and Technology, Shenzhen 518055, China}}
\newcommand{\HFNLHF}{\affiliation{5}{
Hefei National Laboratory, Hefei 230088, China}}
\newcommand{\HFNL}{\affiliation{6}{
Shenzhen Branch, Hefei National Laboratory, Shenzhen 518048, China}}
\newcommand{\NXU}{\affiliation{7}{
School of Physics, Ningxia University, Yinchuan 750021, PR China}}

\begin{document}

\title{Experimental observation of exact quantum critical states}
\date{\today}
\author{Wenhui Huang}
\thanks{These authors contributed equally to this work.}
\affiliation{\SIQSE}\affiliation{\IQA}\affiliation{\GDKL}

\author{Xin-Chi Zhou}
\thanks{These authors contributed equally to this work.}
\affiliation{\PKU}\affiliation{\HFNLHF}

\author{Libo Zhang}
\thanks{These authors contributed equally to this work.}
\affiliation{\SIQSE}\affiliation{\IQA}\affiliation{\GDKL}

\author{Jiawei Zhang}
\thanks{These authors contributed equally to this work.}
\affiliation{\SIQSE}\affiliation{\IQA}\affiliation{\GDKL}

\author{Yuxuan Zhou}
\thanks{These authors contributed equally to this work.}
\affiliation{\IQA}

\author{Bing-Chen Yao}
\thanks{These authors contributed equally to this work.}
\affiliation{\PKU}

\author{Zechen Guo}
\affiliation{\SIQSE}\affiliation{\IQA}\affiliation{\GDKL}

\author{Peisheng Huang}
\affiliation{\NXU}\affiliation{\IQA}

\author{Qixian Li}
\affiliation{\SIQSE}\affiliation{\IQA}\affiliation{\GDKL}

\author{Yongqi Liang}
\affiliation{\SIQSE}\affiliation{\IQA}\affiliation{\GDKL}

\author{Yiting Liu}
\affiliation{\SIQSE}\affiliation{\IQA}\affiliation{\GDKL}

\author{Jiawei Qiu}
\affiliation{\SIQSE}\affiliation{\IQA}\affiliation{\GDKL}

\author{Daxiong Sun}
\affiliation{\SIQSE}\affiliation{\IQA}\affiliation{\GDKL}

\author{Xuandong Sun}
\affiliation{\SIQSE}\affiliation{\IQA}\affiliation{\GDKL}

\author{Zilin Wang}
\affiliation{\NXU}\affiliation{\IQA}

\author{Changrong Xie}
\affiliation{\SIQSE}\affiliation{\IQA}\affiliation{\GDKL}

\author{Yuzhe Xiong}
\affiliation{\SIQSE}\affiliation{\IQA}\affiliation{\GDKL}

\author{Xiaohan Yang}
\affiliation{\SIQSE}\affiliation{\IQA}\affiliation{\GDKL}

\author{Jiajian Zhang}
\affiliation{\SIQSE}\affiliation{\IQA}\affiliation{\GDKL}

\author{Zihao Zhang}
\affiliation{\SIQSE}\affiliation{\IQA}\affiliation{\GDKL}

\author{Ji Chu}
\affiliation{\IQA}

\author{Weijie Guo}
\affiliation{\IQA}

\author{Ji Jiang}
\affiliation{\SIQSE}\affiliation{\IQA}\affiliation{\GDKL}

\author{Xiayu Linpeng}
\affiliation{\IQA}

\author{Wenhui Ren}
\affiliation{\IQA}

\author{Yuefeng Yuan}
\affiliation{\IQA}

\author{Jingjing Niu}
\affiliation{\IQA}\affiliation{\HFNL}

\author{\\Ziyu Tao}
\email{taozy2019@mail.sustech.edu.cn}
\affiliation{\IQA}

\author{Song Liu}
\email{lius3@sustech.edu.cn}
\affiliation{\SIQSE}\affiliation{\IQA}\affiliation{\GDKL}\affiliation{\HFNL}

\author{Youpeng Zhong}
\email{zhongyp@sustech.edu.cn}
\affiliation{\SIQSE}\affiliation{\IQA}\affiliation{\HFNL}

\author{Xiong-Jun Liu}
\email{xiongjunliu@pku.edu.cn}
\affiliation{\PKU}\affiliation{\HFNLHF}\affiliation{\IQA}

\author{Dapeng Yu}
\email{yudapeng@iqasz.cn}
\affiliation{\IQA}\affiliation{\HFNL}

\date{\today}

\begin{abstract}
Anderson localization physics features three fundamental types of eigenstates: extended, localized, and critical, {with the third one exhibiting the exotic properties in-between the former two.}
Confirming the presence of critical states {is challenging, as it typically} necessitates either advancing the analysis to the thermodynamic limit or identifying a universal mechanism which can rigorously determine these states.
Here we report the unambiguous experimental realization of critical states, governed by a rigorous mechanism for exact quantum critical states, and further observe a generalized mechanism that quasiperiodic zeros in hopping couplings protect the critical states.
We implement a programmable quasiperiodic mosaic model with tunable couplings and on-site potentials through a multiple {superconducting qubit quantum system.}
By measuring the time-evolving observables, we identify the coexisting delocalized dynamics and incommensurately distributed zeros in the couplings, which are the defining features of the critical states. We map the localized-to-critical phase transition and demonstrate that critical states persist until quasiperiodic zeros are removed by strong long-range couplings, highlighting a novel generalized mechanism discovered in this experiment and shown with rigorous theory. Finally, we resolve the energy-dependent transition between localized and critical states, revealing the presence of anomalous mobility edges.
\end{abstract}
\maketitle

\begin{figure*}[!tbph]
    \centering
    \includegraphics[width=0.85\textwidth]{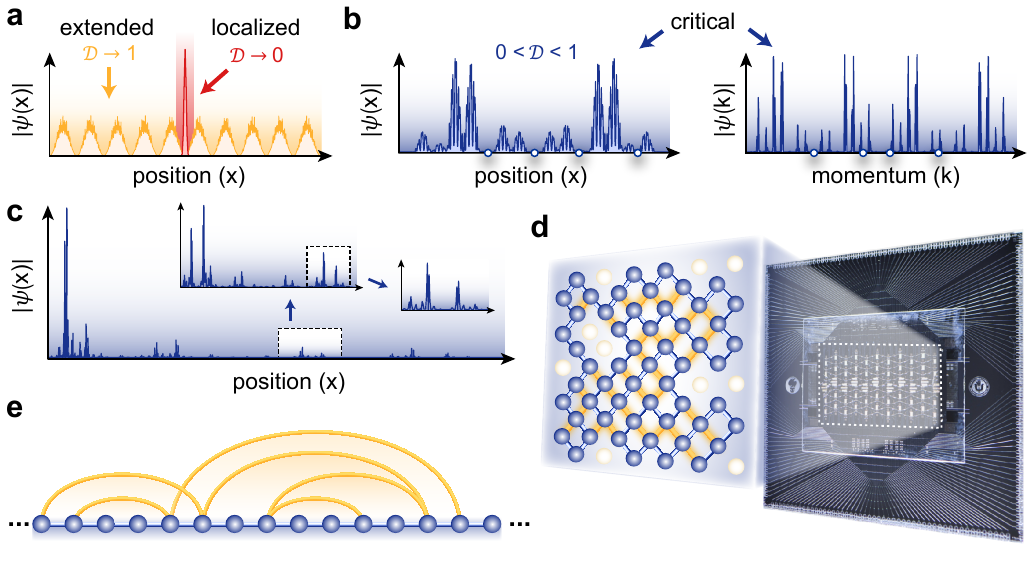}
    \caption{\label{fig1}
    \figtitle{Schematic for the mechanism of quantum critical states.}
    \subfiglabel{a}-\subfiglabel{b} Density profile of the eigenstates in extended, localized (\subfiglabel{a}) and critical phases (\subfiglabel{b}), which are delocalized in both position and momentum space.
    Blue circles denote the nodes caused by incommensurately distributed zeros (IDZs), which are key ingredients for the critical states.
    \subfiglabel{c} Visualization of the self-similar structure characteristic of critical states.
    \subfiglabel{d} The quantum spin model comprising a 2D array of 
    spin qubits.
    Blue spheres represent lattice sites, and blue (orange) bonds mark the nearest-neighbour (long-range) couplings controlled by tunable couplers in the experiment.
    \subfiglabel{e} Illustration of one-dimensional spin chain model incorporating long-range couplings. The 2D geometry 
    facilitates long-range couplings. By activating orange bonds in 2D array and relabeling the system as a 1D chain, nearest-neighbour couplings effectively realize long-range interactions within the redefined 1D configuration.
    }
\end{figure*}

{\em\bf Introduction.}
In an ideal crystal, electrons experience a spatially periodic potential and are characterized by Bloch states that possess lattice translational symmetry, extending uniformly throughout the material.
Introducing disorder breaks the translational symmetry, impeding the extension of the wave functions,
thereby localizing the electronic states; this is known as Anderson localization~\cite{Anderson1958,abrahams1979,lee1985,kramer1993,Evers2008}.
The states at the extended-localized transition, however, are neither fully extended nor localized, but are in a critical phase.
The quantum critical phase is one of the fundamental phases in Anderson localization physics~\cite{hatsugai1990,han1994,wang2016b,wang2021a,wang2022c,liu2022,Zhou2023,goncalves2023b,goncalves2023a},
surpassing the complexity of the extended and localized phases~\cite{
an2018, luschen2018a, kohlert2019a, an2021,wang2022a,gao2024}, and has attracted considerable interest~\cite{liu2024g}.
This phase manifests as delocalized matter waves in both position and momentum spaces with local scale invariance as shown in \rsubfigs{fig1}{a}{c},
reflecting the interplay between self-duality and multifractal structures~\cite{Halsey1986,Ketzmerick1997,Mirlin2006,Dubertrand2014,Yao2019}.
The inclusion of interactions further enriches the physics of the critical phase, with multifractality of the wave function influencing both ground state properties associated with exotic symmetry breaking~\cite{liu2024g,feigelman2007,burmistrov2012,zhao2019,sacepe2020,goncalves2024} and the emergence of non-ergodic many-body critical phases~\cite{wang2020a} at infinite temperature that defies the eigenstate thermalization hypothesis~\cite{dalessio2016, deutsch1991, rigol2008, srednicki1994}.

However, confirming the existence of critical states is significantly challenging.
The localized and extended state wave-functions severely fluctuate when the corresponding localization length or correlation length is comparable to the system size.
Such fluctuations make localized and extended states resemble critical states.
Due to this subtle nature, a precise theoretical framework to characterize {critical} phases has been elusive for decades.
Recent breakthroughs, particularly Avila's global theory, {have refined the rigorous characterization~\cite{simonTraceClassPerturbations1989,jitomirskayaAnalyticQuasiPerodicCocycles2012} of the critical states in quasiperiodic systems~\cite{avila2015,Zhou2023}.}
This establishes a universal mechanism that critical states can emerge when {hopping} couplings in one-dimensional (1D) systems are quasiperiodic and feature incommensurately distributed zeros (IDZs) in the thermodynamic limit~\cite{Zhou2023} as illustrated in \rsubfigsA{fig1}{b}{c}. 
Experimentally, identifying critical states presents comparable difficulties.
Pioneering experimental efforts have been devoted to probing critical phase~\cite{Rispoli2019,goblot2020a,xiao2021,li2023b, shimasaki2024b} in quasiperiodic systems~\cite{suck2002,aubry1980,roati2008,biddle2010a,ganeshan2015,deng2019,yao2019a,wang2020b,longhi2024}.
However, the limited system sizes in experiments introduce severe finite-size effects, causing localized and extended states to exhibit behaviors akin to critical states~\cite{shimasaki2024b}.
This makes the rigorous experimental confirmation of critical states a challenging task.
Despite these obstacles, the elusive observation of transition energies threshold linked to critical states, referred to as anomalous mobility edges (MEs), remains an area ripe for investigation.

In this work,
we report the precise characterization of quantum critical states with smoking-gun evidence and further explore their universal behaviors in a mosaic model~\cite{Zhou2023}
using quantum spin chain formed by superconducting qubits~\cite{Neill2021,Mi2021,Satzinger2021,Mi2021a,Mi2022,Morvan2022,Saxberg2022,
Karamlou2024,Braumueller2021,Karamlou2022,Rosen2024,Rosen2024a,
Ma2019,Du2024,Du2024a,
Gong2021,Chen2021,
Xiang2023,Shi2023,Shi2024,
Yao2023,Zhang2022,Zhang2022a},
which reveal universal rigorous mechanism of such critical states as well as the associated mobility edges.
%
Leveraging the site-resolved control of coupling strengths~\cite{Zhang2022,Zhang2022a,Xu2020},
we first observe the delocalized (localized) dynamics and confirm the presence of {IDZs in the exchange couplings as the key ingredient resulting in the critical phase.}
%
We then turn to a 2D configuration of the coupled qubits, with which we
realize an extended mosaic model with tunable long-range couplings {as illustrated in \rsubfig{fig1}{d}}.
Switching on the long-range couplings,
we uncover a profound generalized mechanism that the IDZs can protect critical states, as long as they are not completely removed {by the long-range terms.} We further examine the elimination of IDZs in the ergodic dynamics and observe the breakdown of localized-critical phase transition when the IDZs disappear, further
confirming the IDZs to be the central ingredient of the universal mechanism 
for stabilizing the critical states.
Finally, we observe the anomalous MEs separating critical from localized states in the spectra by precisely controlling the coupling coefficients and on-site potentials to the exactly solvable point, which match
the results predicted in the thermodynamic limit.
%
{This work sets a standard for the unambiguous experimental detection and characterization of quantum critical states, offering a systematic methodology for their observation, and also presents a tunable quantum platform to further explore the novel physics of the critical states.}

\begin{figure*}[!tbph]
    \centering
    \includegraphics[width=0.95\textwidth]{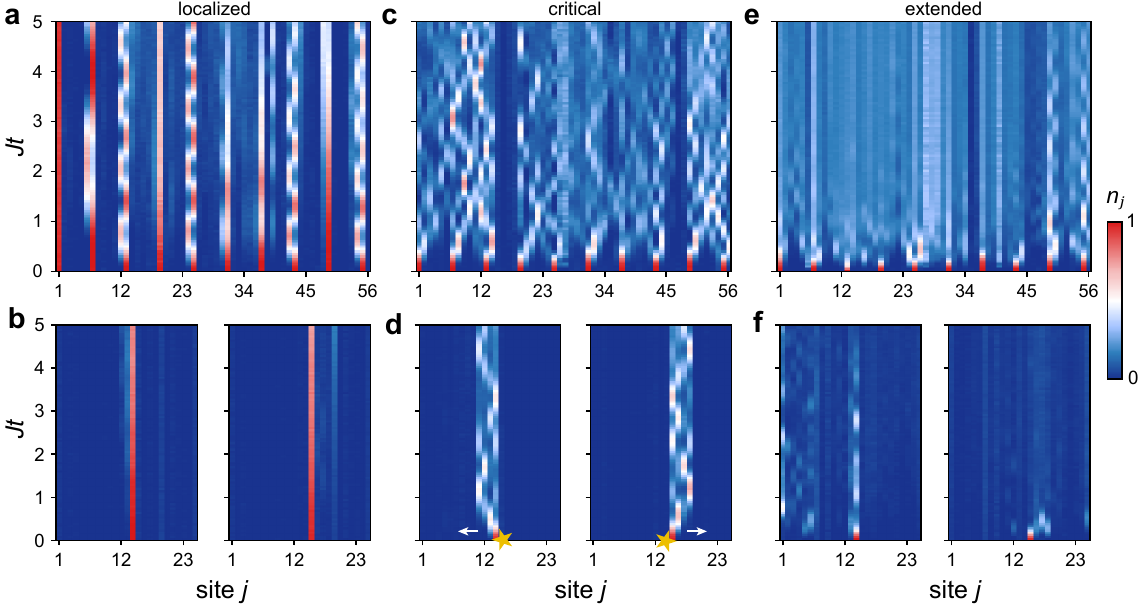}
    \caption{\label{fig2}
    \figtitle{Characteristic dynamics of localized, critical and extended phases.}
    \subfiglabel{a-b}
    Dynamics in the localized phase with $\lambda/J=0.25$.
    \subfiglabel{c-d}
    Dynamics in the critical phase with $\lambda/J=2.5$.
    \subfiglabel{e-f}
    Dynamics in the extended phase with the long-range coupling $J^L_{m,n}=\lambda$ and $\lambda/J=2.5$.
    In \subfiglabel{a},\subfiglabel{c},\subfiglabel{e},
    the top panels show illustrations for the density profiles of each phases,
    and bottom panels show the measured dynamics of on-site population $n_j(t)$,
    where the system is initialized in
    $\vert \psi (t=0)\rangle =  (\prod \sigma^{+}_{6k-5} )\vert 0^{\otimes N}\rangle $,
    with $k$ indexing every sixth site starting from site 1.
    \subfiglabel{b},\subfiglabel{d},\subfiglabel{f}
    show the measured dynamics of $n_j(t)$ with the initial state
    {prepared at the left and right to the zeros in coupling coefficients marked by the stars,}
    highlighting the distinct behaviors of different phases.
    The parameters of systems are
    {$\lambda/(2\pi)=1~\mathrm{MHz}$,} $J^L_{m,n}=0$ (localized phase);
    {$\lambda/(2\pi)=10~\mathrm{MHz}$}, $J^L_{m,n}=0$ (critical phase);
    {$\lambda/(2\pi)=J^L_{m,n}/(2\pi)=10~\mathrm{MHz}$} (extended phase);
    with {$J/(2\pi)=4~\mathrm{MHz}$}, $V_0=0$ and $\theta = \pi/5$.
}
\end{figure*}

{\em\bf Programmable long-range mosaic model.} The 
model studied in the experiment is effectively described by the Hamiltonian of a 1D chain with $N$ qubits 
\begin{eqnarray}
H/\hbar &=& -\sum_{j=1}^{N-1} J_j (\sigma_j^+ \sigma_{j+1}^- + \sigma_j^- \sigma_{j+1}^+) + \sum_{j=1}^{N} V_j \sigma_j^+\sigma_j^- \nonumber \\
&-& \sum_{m,n} J_{m,n}^L (\sigma_m^+ \sigma_{n}^- + \sigma_m^- \sigma_{n}^+),\label{eq:MosaicHam}
\end{eqnarray}
where $\sigma^{+}_{j}$ ($\sigma^{-}_{j}$) denotes the raising (lowering) operator for qubits. {The nearest-neighbour (NN) exchange coupling ($J_j$) has a mosaic quasiperiodic pattern, with the quasiperiodic coupling applied to only the even bonds}, defined as
\begin{equation}
J_j=\left\{\begin{array}{lll}
\lambda, & j=1 & \bmod ~ 2, \\
2 J \cos (2 \pi \alpha j+\theta), & j=0 & \bmod ~ 2.
\end{array}\right.
\end{equation}
Here $\lambda$ and $J$ determine NN coupling strengths, the irrational frequency $\alpha=(\sqrt{5}-1)$/2, and $\theta$ is the phase offset. {The IDZs refer to the vanishing NN couplings $J_{2k}\rightarrow0$ at the bonds with indices $2k$, 
which distribute quasiperiodically over the 1D chain in thermodynamic limit~\cite{Zhou2023}.}
The second term 
is a mosaic on-site potential
\begin{equation}
{V_{j}=2V_{0}\cos\bigr(4\pi\alpha[j/2]+\theta\bigr),} 
\end{equation}
with amplitude $V_0$ and $[j/2]$ taking the integer part. The long-range coupling $J^L_{m,n}$ is introduced by connecting the $m$-th and $n$-th qubits on a 2D lattice configuration (see \rsubfigsA{fig1}{d}{e}).
The long-range coupling enables to explore the mechanism underlying the rigorous critical states and serves as a switch to include extended states in the system.
In the case with $J^L_{m,n}=0$ and $V_0=0$, the system is exactly solvable and exhibits a transition between localized and critical phases at $\lambda=J$.
The localized phase is obtained for $\lambda<J$, with localization length $\xi=2/\log{|J/\lambda|}$, and the critical phase for $\lambda>J$ (see \SeeSupply{}).
The critical states arise from the key mechanism of combining the delocalized nature of the states and the IDZs in NN couplings. {The IDZs cause incommensurate nodes in the delocalized wave-function, which are preserved under dual transformation (\rsubfig{fig1}{b}), rendering the self-dual property, as a characteristic feature of the critical states.}
Turning on long-range couplings may remove the IDZs and drive the critical phase into extended phase.
On the other hand, introducing the on-site potential $V_0>0$ leads to energy-dependent transitions between localized and critical states, where the transition energies $E_{c}$ define the MEs.
Interestingly, along the high-symmetric line $V_{0}=J$, the system becomes exactly solvable again, allowing for an analytical study of localization properties of all states.

In the experiment, all coupling coefficients and on-site potentials are independently tunable, which we utilize to unravel the mechanism of the critical states.
We first demonstrate the fundamental organizing principle for the critical states by toggling the long-range coupling without onsite potential $V_0=0$, highlighting the key role played by the IDZs for realizing the critical states.
Following this, we map out the complete phase diagram for the localized-to-critical phase transition.
Finally, we probe the MEs by adjusting the on-site potential, further showcasing the precise tunability of the system.

\begin{figure*}[th]
    \centering
    \includegraphics[width=0.85\textwidth]{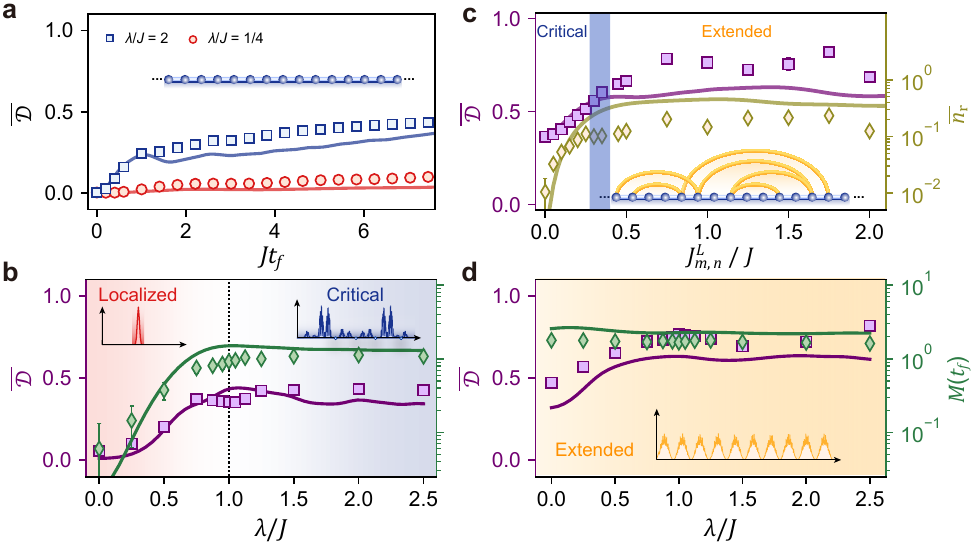}
    \caption{\label{fig_CriLoc_phase}
    \figtitle{{Localized-critical phase transition and its breakdown.}}
    {
    \subfiglabel{a}
    Measured time-averaged observable $\overline{\mathcal{D}}$
    within the time range from $0$ to $t_f$
    for long-ranged coupling $J_{m,n}^L=0$, where $\lambda/J=1/4$ ($\lambda/J=2$) are chosen from a deeply localized (critical) phase.
    \subfiglabel{b}-\subfiglabel{d}
    Characterization of the localized-critical phase transition for $J_{m,n}^L=0$ in \rsubfiglabel{b}
    and its breakdown in the presence of long-ranged coupling in \rsubfiglabel{c-d},
    where \rsubfiglabel{c} shows $\overline{\mathcal{D}}$ and time-averaged $\overline{n}_{\mathrm{r}}=\sum_{j>j_0} n_j$
    for different $J_{m,n}^L/J$ with {$\lambda/(2\pi)=10~\mathrm{MHz}$}, and
    \rsubfiglabel{d} gives $\overline{\mathcal{D}}$ and $M(t_f)$ for different $\lambda/J$ with $J_{m,n}^L=2.5 J$.
    }
    The system is initialized in $\vert \psi (t=0)\rangle =  \vert 1\rangle_{14}$, {$J/(2\pi)=4~\mathrm{MHz}$}, and the size of system is $N=24$. The markers represent experimental data, and the solid lines correspond to simulations.
}
\end{figure*}
{\em\bf Universal mechanism for critical states.}
We first identify the key mechanism relating the critical states to the presence of IDZs by studying the characteristic dynamics in different regimes.
The time evolution of density population
$n_j(t)=\lvert \langle\psi(t)\vert 1\rangle_j \rvert^2$ with $\lvert 1\rangle_j = \sigma^{+}_j \vert 0^{\otimes N} \rangle$,
reveals the unique localization properties of different states as shown in \rfig{fig2}.  {Here $\lvert \psi(t)\rangle=\exp(-iHt) \lvert \psi(0)\rangle$, and $\lvert \psi(0)\rangle$ is the initial state.}
Without the on-site modulation and long-range couplings, we observe only the localized and critical dynamics.
In particular, for the localized phase with $J>\lambda$, the density profile remains confined as shown in \rsubfigsA{fig2}{a}{b}.
{The localization can be understood as follows. Diagonalizing the dominant $J_j=2J\cos(2\pi\alpha j)$ term in Eq.~\eqref{eq:MosaicHam} yields disconnected dimers with quasiperiodically distributed energies, which act as an incommensurate on-site potential.
The smaller uniform coupling $J_j=\lambda$ connects these dimers, and the system resembles double Aubry-Andr{\'e} chain~\cite{aubry1980} in localized phase.}
In comparison, for critical phase with $J<\lambda$ in \rsubfigsA{fig2}{c}{d}, the quantum states propagate while preserving their local configurations, indicating a 
critical behavior. Finally, when the long-range coupling is turned on and $J^L_{m,n}=\lambda$, the density profile rapidly expands over the entire system (see \rsubfigsA{fig2}{e}{f}), erasing the initial configuration and manifesting an extended phase.

The emergence of critical states is observed uniquely connecting to IDZs, as located in bonds at sites $j=14$ and $j=47$ in our sample, through which the tunneling takes longest time (\rsubfig{fig2}{c,d}). 
The nearly vanishing tunneling leads to the unique uni-side quantum dynamics. The lower panel of \rfig{fig2} depicts the single-spin evolution with initial states $\lvert \psi (t=0)\rangle=\lvert 1\rangle_{14}$ and $\lvert 1\rangle_{15}$, respectively.
{For critical phase, we observe that the wave-packet propagates in one side of the bond connecting sites $j=14$ and $15$, with negligible probability in another side (see \rsubfig{fig2}{d}). This uni-side propagation quantum dynamics provides a characteristic signature to identify and benchmark IDZs within the system.} 
In contrast, for the long-range coupling $J^L_{m,n}=\lambda$, the uni-side quantum dynamics breaks down, with the propagation across the bond at $j=14$ being clearly observed (\rsubfig{fig2}{f}). Thus the IDZs disappear and the system turns to extended phase (\rsubfig{fig2}{e}). 
The distinct quantum dynamics associated with the IDZs enables a comprehensive exploration of the entire phase diagram based on the present programmable 2D sample by tuning the long-range couplings.

{\em\bf Localized-critical-extended transitions.}
We then investigate the phase transition between localized and critical states by mapping out the phase diagram for the mosaic model with $V_0=0$.
To characterize the transition {and to avoid fast oscillation (see \SeeSupply{})}, we employ the time-averaged {fractal dimension} 
$\overline{\mathcal{D}}=(1/t_f)\int_{0}^{t_f}[\mathcal{D}(\tau)-\mathcal{D}(0)]d\tau$ and the integrated width $M(t_f)$ within the time range from $0$ to $t_f$.
{Here the observable $\mathcal{D}=-\log\sum_{j}|u_{m,j}|^{4}/\log{N}$ is the fractal dimension of the state $\lvert\psi_{m}\rangle=\sum_{j=1}^{N}u_{m,j}\sigma_{j}^{+}\lvert 0^{\otimes N}\rangle$, and quantifies the effective spatial dimension of the state. 
}
Given that $\mathcal{D}$ is always between 0 and 1 for the finite system size,
we also introduce the integrated width $M(t_f)=(1/t_f)\int_{0}^{t_f}[W(\tau)-W(0)]d\tau$, with $W(\tau)=\sum_{j}\sqrt{|j-j_{0}|}\langle n_{j}(\tau)\rangle$,
to characterize the expansion of the spin transport {initially located at $j_0$} during early-time dynamics and serve as a complementary metric.
%
We conduct the experiment on a 24-site subarray, which is sufficient to capture the dynamics near the IDZs within the relevant time scale as shown in \rsubfigs{fig2}{d}{f}.
For instance, we present
the time-averaged observable $\overline{\mathcal{D}}$ within the time range from $0$ to $t_f$
for typical critical (localized) phases as $\lambda/J=2$ ($\lambda/J=1/4$) in \rsubfig{fig_CriLoc_phase}{a}, by initializing the system in the state $\lvert \psi (t=0)\rangle=\lvert 1\rangle_{14}$ located near the IDZ, which exhibits the most distinct behaviors in \rfig{fig2}.
We then benchmark the phase diagram of the mosaic model in \rsubfig{fig_CriLoc_phase}{b}.
By varying $\lambda/J$ from $0$ to $2.5$ while maintaining {$J/(2\pi)=4~\mathrm{MHz}$} and $t_f=\mathrm{300~ns}$,
the system transitions from a deeply localized phase to a critical phase.
This transition is evidenced by the increase of $\overline{\mathcal{D}}$ from nearly zero to a saturated plateau about $0.4$, and the corresponding increase of $M(t_f)$ from vanishingly small to a finite value.
The nearly simultaneous saturation of $\overline{\mathcal{D}}$ and $M(t_f)$ signals a phase transition between localized ($\lambda<J$) and critical phase ($\lambda>J$) near $\lambda/J=1$, matching well the theoretical calculations. 

\begin{figure*}[th]
    \centering
    \includegraphics[width=0.99\textwidth]{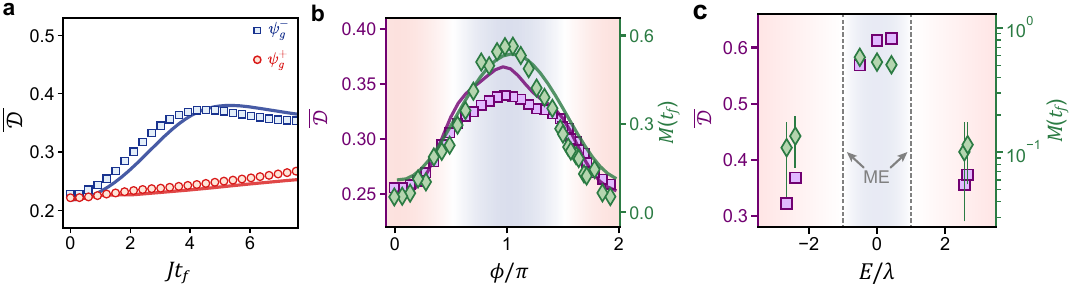}
    \caption{\label{fig_ME}
    \figtitle{Mobility edges in the quasiperiodic mosaic lattice.}
    \subfiglabel{a}
    Measured time-averaged observable $\overline{\mathcal{D}}$,
    where the system is quenched from the localized state $\vert \psi_g^+\rangle$ and zero-energy state $\vert \psi_g^-\rangle$ of the Hamiltonian $H$ as $\lambda \to 0$ with $\vert \psi_g^{\pm}\rangle$ $= (\vert 1\rangle_{12}\pm\vert 1 \rangle_{13})/\sqrt{2}$.
    The site $j$ ranging from 1 to $N=24$ and $\theta=\pi/5$.
    \subfiglabel{b}-\subfiglabel{c}
    Measured $\overline{\mathcal{D}}$ and $M(t_f)$ for the system quenched from
    the superposition of localized and critical state $\vert \psi^{\phi}_n \rangle = (\vert 1\rangle_{12} + e^{i\phi}\vert 1 \rangle_{13})/\sqrt{2}$ (\subfiglabel{b}),
    as well as for states $\vert \psi^{\pm}_n \rangle$ located
    near the edge (inside the novel mobility edges) and center (outside the novel mobility edges) of the spectrum (\subfiglabel{c}), where $t_f=\mathrm{300~ns}$ and $\vert \psi^{\pm}_n \rangle = (\vert 1\rangle_{n}\pm\vert 1 \rangle_{n+1})/\sqrt{2}$. The dashed lines mark the mobility edges separating the localized and critical states.
    The markers represent experimental data, and the solid lines correspond to numerical simulations.
}
\end{figure*}

With the long-range couplings, we experimentally discover a novel generalized mechanism that IDZs and critical states are robust to weak $J_{m,n}^L$.
The transition from critical to extended states occurs only when $J_{m,n}^L$ exceeds a threshold, and can be observed by measuring the characteristic uni-side propagation dynamics of critical states. We quantify the observation through the time-averaged density population across IDZ (say at $j=j_0$) observed in above section $\overline{n}_{\mathrm{r}}=\sum_{j>j_0} n_j$.
The critical state turns to an extended one when both $\overline{n}_r$ and $\overline{\mathcal{D}}$ saturate, indicating that the uni-side dynamics break down and IDZs fully disappear.
As shown in \rsubfig{fig_CriLoc_phase}{c}, the small long-range coupling does not ruin the critical states, with $\overline{n}_r\approx0$. Instead, the IDZs of the dominating $J_j$-terms 
are dressed by long-range couplings, and stabilize the critical states in this regime. We observe that when $J_{m,n}^L/J\gtrsim 0.4$, both $\overline{\mathcal{D}}$ and $\overline{n}_r$ saturate as marked by the blue shadow, consistent with the numerical results. 
{Note that systematic deviations are found between experiment and theory in the extended regime (\rsubfig{fig_CriLoc_phase}{c}) and is due to the experimental imperfections (e.g. stray diagonal couplings~\cite{Andersen2025}), involving which in numerical simulation can remove the deviations (see Suppementary Information).}
In theory, we first model the dressed IDZs due to the next-nearest-neighbor (NNN) couplings $J_{\rm nn}$, and show the critical-extended phase transition analytically by renormalization group method~\cite{goncalves2023b,goncalves2023a} (see Supplementary Material). The transition criterion is
\begin{equation}
J_{\mathrm{nn}}>\max\big(J,\sqrt{J\lambda}\big),
\end{equation}
indicating that critical states are robustly protected by IDZs and critical-extended transition only happens only when NNN couplings exceed a certain threshold. Furthermore, the transition $J_{\rm nn}$ decreases when next-next-nearest-neighbor hopping $\mu$ is involved. When $\mu$ is relatively small,
the transition criterion becomes
\begin{equation}\label{transitionpoint}
    J_{\rm nn}>\sqrt{J \max\left(J, \lambda, \mu\right)- \lambda \mu}.
\end{equation}
For $\lambda=2.5J$, one finds that the transition {$J^c_{\rm nn}\approx0.4J$} requires to take {$\mu\approx0.93J$.} 
While the Eq.~\eqref{transitionpoint} is more precise for relatively small $\mu$, this result accounts for the experimental observation that the critical-extended transition occurs at moderate $J_{m,n}^L/J$ due to long-range couplings. {Moreover, through a finite-size analysis based on the full Hamiltonian Eq.~\eqref{eq:MosaicHam}, we obtain the critical-extended transition at $J_{m,n}^L\approx0.3J$, close to the experimental observation (see Supplementary Information).}  

To further illustrate the transition between critical and extended phases, we also measure the phase diagram versus $\lambda$ with fixed long-range coupling $J_{m,n}^L$, as shown in \rsubfig{fig_CriLoc_phase}{d}.
Turning on a large uniform long-range coupling $J_{m,n}^L=2.5J$ always drives the system at any $\lambda$ into an extended phase, 
while increasing $\lambda$ further delocalizes the system, resulting in larger accumulated $\overline{\mathcal{D}}$ and wider $M(t_f)$ compared to the case $\lambda=0$. This feature further confirms the transition from critical to extended phase under large $J_{m,n}$. 
These measurements show the rigorous criteria for quantum critical states: the IDZs in the dominating hopping couplings $J_j$ and the delocalized nature in the regime $\lambda>J$.
The two criteria guarantee the self-similar and scale-invariance of critical states.

%

{\em\bf Probing mobility edges.}
Having established the experimental detection and characterization of critical states, we can probe the {existence of} new MEs between critical and localized states~\cite{Zhou2023}. For this we turn off long-range couplings but switch on the quasiperiodic potential $V_j$. 
From numerical results, we show when $V_{0}/J$ is small, only the edges of the spectrum remain localized, while for $V_{0}\gg J$, most of the spectrum becomes localized (see Supplementary Material). At the high symmetry line $V_{0}=J$, the system is again exactly solvable, yielding MEs at $E_c=\pm \lambda$, with critical (localized) states for the region $\vert E\vert < \vert \lambda \vert$ ($\vert E\vert > \vert \lambda \vert$).
In particular, in this regime the ratio of localized to critical states is $1:1$, which facilitates to probe the physics of the anomalous MEs. 

%
We 
{demonstrate the existence of} MEs for $V_{0}/J=1$ by setting {$J/(2\pi)=4~{\rm MHz}$}. 
We start by 
{exploring} the ME enriched quench dynamics of entangled states near the IDZ. Specifically, we initialize the system in $\vert \psi_g^{\pm}\rangle = (\vert 1\rangle_{12}\pm\vert 1 \rangle_{13})/\sqrt{2}$,
which are eigenstates of the Hamiltonian for $\lambda\to 0$.
%
We then evolve the initial states under the Hamiltonian at $\lambda/J=1/4$. 
In \rsubfig{fig_ME}{a},
$\overline{\mathcal{D}}$ of the initially localized state $\vert\psi^+_g\rangle$ ($\vert\psi^-_g\rangle$) displays a slow (fast) accumulation. Such qualitatively different dynamics of $\vert \psi_g^{\pm}\rangle$ within the same parameter of the system, indicating the presence of the ME.
We further incorporate more initial states with energies by adjusting the relative phase $\phi$ of the state $\lvert\psi^{\phi}_g\rangle = (\vert 1\rangle_{12} + e^{i\phi}\vert 1 \rangle_{13})/\sqrt{2}$.
As shown in \rsubfig{fig_ME}{b}, $M(t_f)$ displays a non-trivial pattern of rise-fall as the $\phi$ evolves,
and $\overline{\mathcal{D}}$ shows a similar trend. 
This indicates that the changes in $\phi$ alter the ratio of localized to critical states, leading to distinct dynamics and suggesting the coexistence of these states within the spectrum resulted by the MEs.
{The deviation between experimental observation and numerical simulation near $\phi/\pi = 1$ is attributed to the fact that the states around this region are critical and are more significantly affected by the spatially distributed experimental imperfections (see Supplementary Information for details).}

Finally, we probe the localized-to-critical transition behavior by tuning the energy of initial state to cross the ME: $\vert E \vert =\lambda$, with $\lambda/J=1.5$. The initial states $\lvert\psi^{\pm}_n \rangle= (\vert 1\rangle_{n}\pm\vert 1 \rangle_{n+1})/\sqrt{2}$ for various $n$, as shown in \rsubfig{fig_ME}{c}, 
are chosen to realize typical energies both inside and outside the MEs, with $n=\{4,8,10,12,16,18\}$ for the plus sign and an additional $n=18$ for the minus sign.
For initial states prepared with $|E|>\lambda$, we observe that the $\overline{\mathcal{D}}$ and $M(t_f)$ saturates to small values, implying the major contribution from localized orbitals.
Conversely, if the initial state is prepared near the center of the spectrum ($|E|<\lambda$), its dynamics displays clear critical characteristics with much larger $M(t_f)$ accompanied with $\overline{\mathcal{D}}\sim0.62$.
{We note that to determine the precise energy of the ME needs to prepare states sufficiently close to $E=\lambda$, which remains challenging for the current experimental setting (see Supplementary Information). However, here the observed energy-dependent spin dynamics spreading confirms the presence of the ME between the localized and critical orbitals.}

{\em\bf Discussion and outlook.}
We have experimentally {realized and studied} quantum critical states in a superconducting quantum simulator with tunable (long-range) couplings and on-site potentials. We discovered in experiment and showed with rigorous theory a novel generalized mechanism of critical states, highlighting the role of IDZs in stabilizing the critical state. In particular, the critical states survive as long as the quasiperiodic zeros are not fully removed by long-range couplings. We further investigate the MEs between critical and localized states, revealing enriched spin evolution and energy-dependent quench dynamics near the IDZs.

This study established a standard to rigorously characterize the critical states in experiment, applicable to broad range of 
quantum simulation platforms including trapped ions and neutral atoms as well,
and provided a versatile framework 
to further explore novel quantum physics of critical states, in particular for higher dimensional systems and the many-body regime. For instance, the inter-particle interactions can be introduced by incorporating anharmonicity in the qubits~\cite{tao}, and shall lead to the emergence of novel phases that have not been precisely observed before.
The multifractal nature of critical states renders the ground phase of the system highly sensitive to interactions, enriching the exotic symmetry-breaking phases~\cite{liu2024g,feigelman2007,burmistrov2012,zhao2019,sacepe2020,goncalves2024}.
For higher energy states, the non-interacting critical states turn to many-body critical phases~\cite{wang2020a}---a third type of fundamental phase in disordered system, which is different from the thermal phase and many-body localization. Building on the present setting, these highly important quantum many-body phases hold the promise for our next studies.

Beyond the closed system framework, our 2D tunable array 
offers high controllability of noise~\cite{Tao2024,liang2024}, enabling the in-depth study of the robustness of critical states and anomalous MEs against decoherence.
Also, by integrating 
tailored dissipation, one can explore dissipation-induced critical phases and mobility edges~\cite{longhi2024a,liu2024j}. {Moreover, the combination of many-body interactions and controlled noise opens up the way to experimentally probe transitions between many-body critical phase in open systems and generic thermalized states. This is an area that remains unexplored both theoretically and experimentally, yet is of fundamental importance.}


%

\bibliographystyle{ref_style}

\vspace{10pt}
\noindent
{\bf  Data availability}

\noindent
 The data that support the plots within this paper and other findings of this study are available from the corresponding authors upon request.

\vspace{10pt}
\noindent
{\bf  Code availability}

\noindent
All the relevant source codes are available from the corresponding authors upon request.

\vspace{10pt}
\noindent
{\bf  Acknowledgements}
\noindent
{\small We thank Qi Zhou and Yucheng Wang for helpful discussions. This work was supported by Quantum Science and Technology-National Science and Technology Major Project (No.2021ZD0301703 and No.2021ZD0302000), National Key Research and Development Program of China (2021YFA1400900), the Science, Technology and Innovation Commission of Shenzhen Municipality (KQTD20210811090049034), the National Natural Science Foundation of China (No.12174178, No.12425401 and No.12261160368), and Shanghai Municipal Science and Technology Major Project (No.2019SHZDZX01)
}

\vspace{10pt}
\noindent
{\bf  Author contributions}

\noindent
W.H. designed and tested the device. Libo Z. and Yuxuan Z. fabricated the device under the supervision of S.L.
Z.T. collected the data, and analyzed the data with help of X.-C.Z.. X.-C.Z. and B.-C.Y. provided theoretical and numerical studies under the supervision of X.-J.L.. Jiawei Z. built the microwave electronics. X.-J.L. conceived the project. Youpeng Z., X.-J.L., and D.Y. supervised the project. All authors contributed to the discussion and preparation of the manuscript.

\noindent
{\small
}

\vspace{10pt}
\noindent
{\bf  Competing interests}

\noindent
The authors declare no competing interests.



\section*{Supplementary material}

# Supplementary information for "Exact quantum critical states with a superconducting quantum processor"

**CONTENTS**

# I. THEORETICAL DETAILS ON THE MOSAIC LATTICE MODEL

In the main text, we simulate the mosaic lattice model with and without on-site potential on a superconducting quantum processor and demonstrate its dynamics associated with the density profile of eigenstates. Here, we provide further details regarding the analytic results of the models.

## A. Review of global theory of one-frequency cocycle

We begin by reviewing the global theory characterizing the one-dimensional (1D) quasiperiodic systems and applying it to the model investigated in this work. The global theory analyzes the one-frequency SL(2) cocycle, which describes systems with nearest-neighbor hopping and a single incommensurate factor. A well-known example is the Aubry-André-Harper (AAH) model [S1], or the almost-Mathieu operator in mathematics [S2]. Eigenstates in such quasiperiodic systems can be categorized as extended, localized, and critical, which belong to the absolutely continuous, pure point, and singularly continuous spectra, respectively.

The localization properties of the eigenstates can be fully characterized by examining the Lyapunov exponent (LE), denoted as $\gamma$, using Avila's global theory [S2]. Consider a generic 1D quasiperiodic system with one-frequency quasiperiodic modulation, with the Hamiltonian given by

$$H/\hbar = \sum_j J_j \left(\sigma_j^+ \sigma_{j+1}^- + \sigma_j^- \sigma_{j+1}^+\right) + \sum_j V_j(\alpha)\sigma_j^+ \sigma_j^-. \tag{S1}$$

The one-frequency quasiperiodic modulation for such Hamiltonian means that the hopping coefficients $J_j$ and on-site potential $V_j$ can be modulated by a single frequency, characterized by an irrational number $\alpha$. For an eigenstate $|\Psi\rangle$ with the energy $E$, expanded over the real-space basis as $|\Psi\rangle = \sum_j u_j \sigma_j^+ |0^{\otimes N}\rangle$, the Schrödinger equation $H|\Psi\rangle = E|\Psi\rangle$ leads to:

$$J_{j-1}u_{j-1} + J_j u_{j+1} + V_j u_j = E u_j. \tag{S2}$$

The LE for the eigenenergy $E$, denoted by $\gamma_0(E)$, is derived by considering:

$$\gamma_\epsilon(E) = \lim_{n\to\infty} \frac{1}{2\pi n} \int \ln \left\| \prod_{j=1}^n T_j(\theta + i\epsilon) \right\| d\theta, \tag{S3}$$

where $||A||$ is the norm of the matrix A, $\epsilon$ is the imaginary part of the complexified $\theta$ and $T_j$ is the transfer matrix at site $j$ satisfying $\left( u_{j+1}, u_j \right)^\mathsf{T} = T_j \left( u_j, u_{j-1} \right)^\mathsf{T}$, with

$$T_j = \begin{pmatrix} \frac{E-V_j}{J_j} & -\frac{J_{j-1}}{J_j} \\ 1 & 0 \end{pmatrix}. \tag{S4}$$

The key result from Avila's global theory is that $\gamma_\epsilon(E)$, as a function of $\epsilon$, is convex, continuous and piecewise linear with a quantized *acceleration* $\omega(E;\epsilon)$:

$$\omega(E;\epsilon) = \lim_{\epsilon \to 0^+} \frac{1}{2\pi\epsilon}[\gamma_\epsilon(E) - \gamma_0(E)] = \mathbb{Z}. \tag{S5}$$

The non-negative LE determines the localization properties of the eigenstates: $\gamma(E) > 0$ implies the state with the energy $E$ belongs to the pure point spectrum and is localized with the localization length $\xi(E) = \gamma^{-1}$. Conversely, $\gamma(E) = 0$ indicates the state is delocalized, possessing an infinite localization length. Delocalized states can be further categorized into extended (absolutely continuous spectrum) and critical states (singularly continuous spectrum) [S3]. Critical states, corresponding to the singularly continuous spectrum, can be realized by imposing one of two fundamental conditions on the delocalized states ($\gamma = 0$): either introducing an unbounded quasiperiodic on-site potential or incorporating incommensurately distributed zeros (IDZs) in the hopping terms [S4]. The latter mechanism is the primary focus of our experiment, as the former is generally unfeasible in physical systems due to the requirement of divergent on-site energies. Both conditions effectively partition the 1D system into multiple subchains, prompting the delocalized orbitals to reorganize within these subchains and thereby giving rise to critical states.

### B. Generalized mosaic lattice model in the uniform potential limit

We analytically characterize the model in the main text for the limit of nearest-neighbor coupling (i.e., long-range coupling $J_{m,n}^L = 0$) and uniform potential $V_0 = 0$. The Hamiltonian is

$$H/\hbar = \sum_j J_j \left( \sigma_j^+ \sigma_{j+1}^- + \sigma_j^- \sigma_{j+1}^+ \right), \tag{S6}$$

with

$$J_j = \begin{cases} \lambda & j = 1 \mod 2, \\ 2J\cos(2\pi\alpha j + \theta) & j = 0 \mod 2. \end{cases} \tag{S7}$$

3

The coupling coefficients exhibit a mosaic pattern, thus we consider the two-fold transfer matrix

$$T_j = \begin{pmatrix} \frac{E}{2J\cos(2\pi\alpha j + \theta)} & -\frac{\lambda}{2J\cos(2\pi\alpha j + \theta)} \\ 1 & 0 \end{pmatrix} \begin{pmatrix} \frac{E}{\lambda} & -\frac{2J\cos(2\pi\alpha j + \theta)}{\lambda} \\ 1 & 0 \end{pmatrix},$$

$$= \frac{1}{2\cos(2\pi\alpha j + \theta)} \begin{pmatrix} \frac{E^2 - 4J^2\cos^2(2\pi\alpha j + \theta)}{J\lambda} & -E/J \\ E/J & -\lambda/J \end{pmatrix},$$

$$= \frac{1}{2\cos(2\pi\alpha j + \theta)} \tilde{T}_j. \tag{S8}$$

By complexifying the phase of the system as $\theta \to \theta + i\epsilon$ and taking the limit $\epsilon \to \infty$, we calculate the LE using Eq. S3. The first term can be obtained using Jenson's formula

$$\int_0^{2\pi} \ln\left|\frac{1}{2\cos(\theta + i\epsilon)}\right| d\theta = -2\pi|\epsilon|. \tag{S9}$$

The second term is

$$\tilde{\gamma}_\epsilon(E) = \lim_{n\to\infty} \frac{1}{2\pi n} \int \ln\left\|\prod_{j=1}^n \tilde{T}_j(\theta + i\epsilon)\right\| d\theta,$$

$$= \lim_{n\to\infty} \frac{1}{2\pi n} \int \ln\left\|\prod_{j=1}^n (J/\lambda) e^{i4\pi\alpha j} e^{i2\theta} e^{2\epsilon}\right\| d\theta,$$

$$= \ln|J/\lambda| + 2|\epsilon|. \tag{S10}$$

Thus we obtain

$$2\gamma_\epsilon = \ln|J/\lambda| + |\epsilon|, \tag{S11}$$

in which the factor 2 arises from the counting of the two-step transfer matrix. According to global theory [S2], $\gamma_\epsilon$ as a function of $\epsilon$ is a convex, piecewise linear function with quantized slopes with respect to $\epsilon$. And it is an affine function in the neighborhood of $\epsilon = 0$. Consequently, the LE for the system is

$$2\gamma_0 = \max\{\ln|J/\lambda|, 0\}. \tag{S12}$$

For $|\lambda| < |J|$, the system is in the localized phase with the localization length given by

$$\xi = \frac{1}{\gamma_0} = \frac{2}{\ln|J/\lambda|}. \tag{S13}$$

For $|\lambda| > |J|$, the $\gamma_0 = 0$, and the system exhibits a sequence of site indices $\{j_k\}$ where the coupling coefficients vanish in the thermodynamic limit ($J_{j_k} \to 0$), which are the IDZs of the hopping terms. Thus, the system is in the critical phase [S3, S4].

4

We further numerically compute the phase diagram of the model by directly diagonalizing the Hamiltonian and calculate the fractal dimension (FD) for the $m$th eigenstate $|\psi\rangle = \sum_{j=1}^{L} u_{m,j} \sigma_j^+ |0^{\otimes N}\rangle$, which is defined as $\mathrm{FD} = -\lim_{L\to\infty} \ln(\mathrm{IPR})/\ln(L)$, with the inverse partition ratio given by $\mathrm{IPR} = \sum_j |u_{m,j}|^4$. For localized states, FD approaches 0, while for extended states, FD approaches 1. Critical states exhibit FD values between 0 and 1. Fig. S1 shows FD as a function of energy $E$ and $\lambda/J$, with transitions between localized and critical phases observed, consistent with our analytical results. The FD approaches 0 for $|\lambda| < |J|$, indicating localization, and increases towards a critical regime for $|\lambda| > |J|$. In addition, for the finite system we consider here, the FD is not exactly zero but remains close to zero for the localized states. One can observe that the FD is closer to 0 for smaller $\lambda/J$ within the localized phase, aligning with the analytical result that a smaller $\lambda/J$ corresponds to a shorter localization length and more localized compared to larger $\lambda/J$.

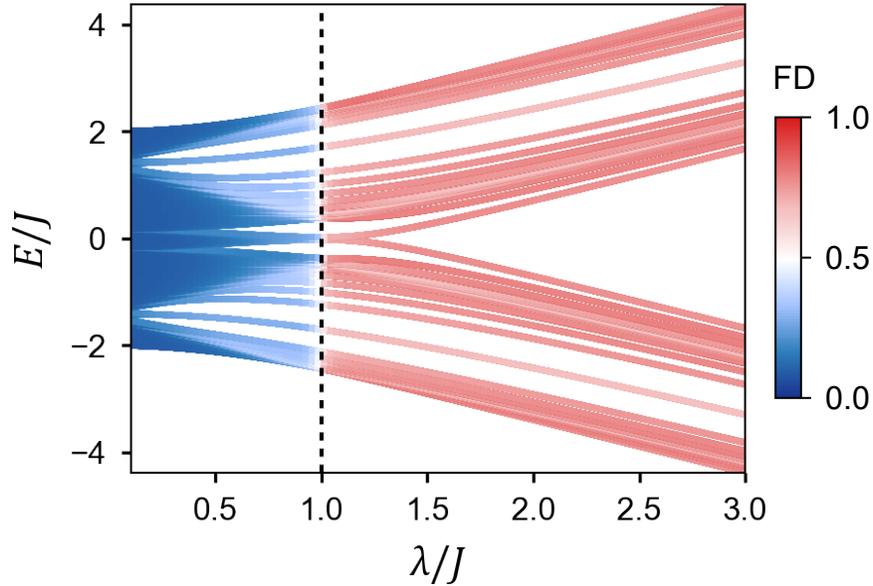

Supplementary Fig. S1. **Phase diagram of generalized mosaic model with $J_{m,n}^L = V_0 = 0$.** Fractal dimension (FD) of the eigenstate as a function of energy $E$ and $\lambda/J$ for a system size of $L = 2584$. The transition from localized to critical phases occurs as $\lambda/J$ increases, consistent with the analytic results.

## C. Generalized mosaic lattice model along high symmetry lines

We next provide an exact characterization of the model with a mosaic incommensurate on-site potential, while maintaining $J^L_{m,n} = 0$. The Hamiltonian is

$$H/\hbar = \sum_j J_j \left(\sigma_j^+ \sigma_{j+1}^- + \sigma_j^- \sigma_{j+1}^+\right) + \sum_j V_j \sigma_j^+ \sigma_j^-, \tag{S14}$$

where the definition of $J_j$ remains the same as before, and the on-site potential $V_j$ is given by

$$V_j = \begin{cases} 2V_0 \cos[2\pi\alpha(j-1) + \theta] & j = 1 \mod 2, \\ 2V_0 \cos(2\pi\alpha j + \theta) & j = 0 \mod 2. \end{cases} \tag{S15}$$

Introducing the on-site potential generically breaks the solvability of the system; however, it remains analytically tractable along high-symmetry lines where $|V_0| = |J|$. Here we focus on the case $V_0 = J > 0$ without loss of generality. The two-step transfer matrix now becomes

$$\begin{aligned} T_j &= \frac{1}{M} \begin{pmatrix} (E^2 - 2EJM)/J\lambda & -(E/J - M) \\ E/J - M & -\lambda/J \end{pmatrix}, \\ &= \frac{1}{2\cos(2\pi\alpha j + \theta)} \tilde{T}_j, \end{aligned} \tag{S16}$$

with $M = 2\cos(2\pi\alpha j + \theta)$. By complexifying the phase $\theta \to \theta + i\epsilon$ and taking the limit $\epsilon \to \infty$, we calculate the LE as follows:

$$\begin{aligned} 2\gamma_\epsilon &= \frac{1}{2\pi} \int_0^{2\pi} \ln\left|\frac{1}{2\cos(2\pi\alpha j + \theta)}\right| d\theta + \lim_{n\to\infty} \frac{1}{2\pi n} \int \ln\left\|\prod_{j=1}^n T_j(\theta + i\epsilon)\right\| d\theta, \\ &= -|\epsilon| + \lim_{n\to\infty} \frac{1}{2\pi n} \int \ln\left\|\prod_{j=1}^n \begin{pmatrix} -2E/\lambda & 1 \\ -1 & 0 \end{pmatrix} e^{-i(2\pi a\alpha j + \theta)} e^\epsilon \right\| d\theta, \\ &= \ln\left||E/\lambda| + \sqrt{E^2/\lambda^2 - 1}\right|. \end{aligned} \tag{S17}$$

Following the same reasoning as in the previous section, the LE for the system is given by

$$2\gamma_0 = \max\left\{\ln\left||E/\lambda| + \sqrt{E^2/\lambda^2 - 1}\right|, 0\right\}. \tag{S18}$$

The critical energies for the LE transition from zero to a non-zero value, known as the mobility edges (MEs), are

$$E_c = \pm\lambda. \tag{S19}$$

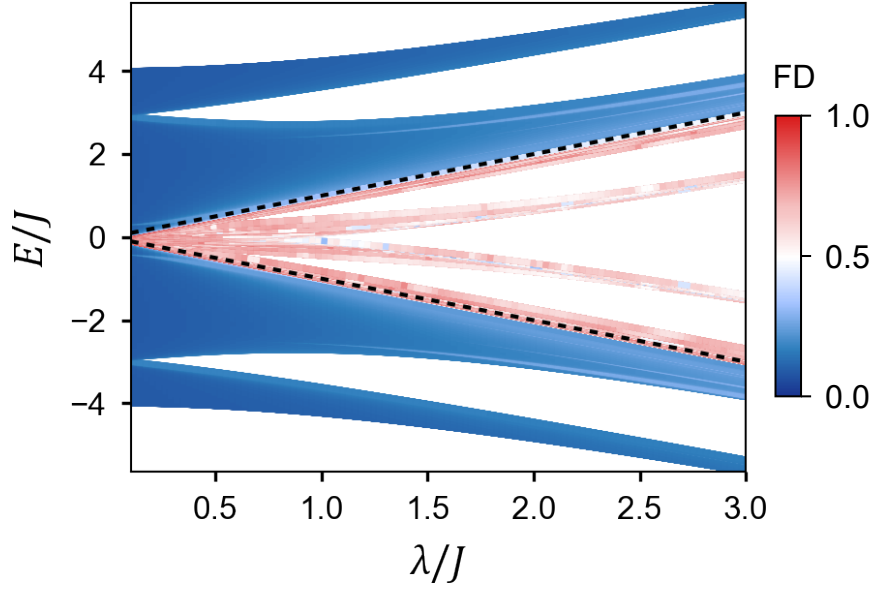

Supplementary Fig. S2. **Phase diagram of generalized mosaic model with $J_{m,n}^L = 0$ and $V_0 = J$.** The fractal dimension (FD) of the eigenstate with the energy $E$ for a system size of $L = 2584$. The dashed lines mark the positions of mobility edges at $E = \pm\lambda$. Eigenstates are localized for energies $|E| > |\lambda|$ and critical for energies $|E| < |\lambda|$, as predicted by the analytical results.

For eigenenergies $|E| > |\lambda|$, the corresponding eigenstates are localized with the localization length given by

$$\xi(E) = \frac{1}{\gamma_0} = \frac{2}{\ln\left||E/\lambda| + \sqrt{E^2/\lambda^2 - 1}\right|}. \tag{S20}$$

While for the eigenenergies $|E| < |\lambda|$, the $\gamma = 0$, and the system exhibits IDZs in the thermodynamic limit. Consequently, the corresponding eigenstates are critical states.

Fig. S2 displays the FD of the eigenstates as a function of energy $E$ for different $\lambda/J$. The dashed lines indicate the positions of the MEs at $E = \pm\lambda$. FD approaches 0 for the eigenstates with $|E| > |\lambda|$, confirming their localized nature, and takes values between 0 and 1 for eigenstates with $|E| < |\lambda|$, consistent with critical states. Additionally, for the finite system size considered ($L = 2584$), the FD does not reach exactly zero but remains close to zero in the localized phase. FD is closer to 0 for eigenstates with energies distant from the MEs. This observation aligns with analytical predictions: eigenstates located far from the MEs exhibit shorter localization lengths, resulting in lower FD values. As eigenstates approach the MEs, the localization length increases and diverges at the MEs, leading to

higher FD values.

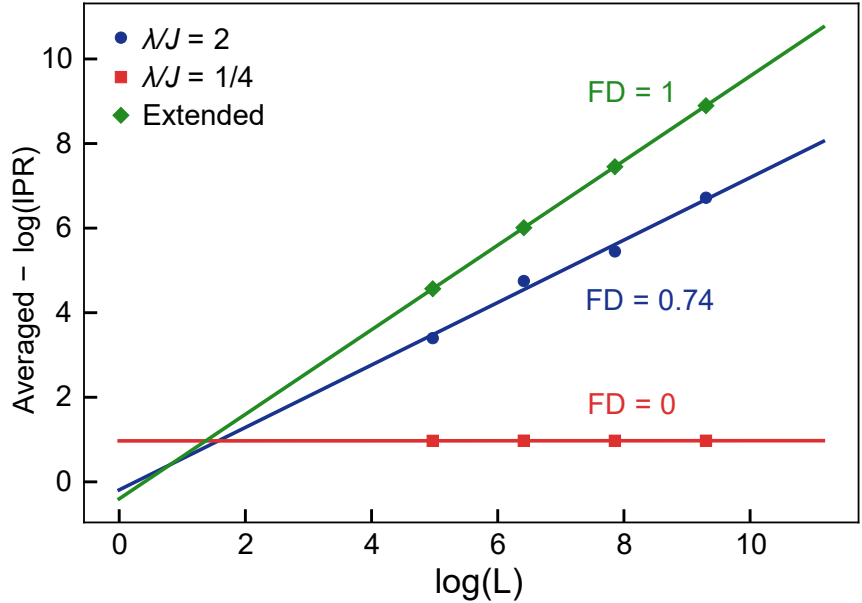

Supplementary Fig. S3.  **Finite size scaling analysis of the fractal dimension.** Plot of the averaged $-\log(\text{IPR})$ of all eigenstates versus $\log(L)$ for system size ranging from $L = 144$ to $L = 10946$. Solid lines represent linear fits, with their slopes corresponding to the fractal dimensions. The fitted FD values are zero in the localized phase and approximately 0.74 in the critical phase for this model. And the fitted FD is 1 in the extended phase.

### D. Finite size scaling for generalized mosaic lattice model

We investigate the finite size scaling of fractal dimension for the eigenstates in the generalized mosaic lattice model without long-range coupling, i.e., $J_{m,n}^L = 0$, in the presence and absence of a quasiperiodic on-site potential. We numerically perform the finite size scaling from system sizes $L = 144$ to $L = 10946$ to evaluate the fractal dimension of the model in the thermodynamic limit $L \to \infty$. This involves calculating the inverse partition ratio IPR for each finite size system and then analyzing the scaling of IPR as a function of $L$. The slope of the logarithmic plot, $-\log(\text{IPR})$ versus $\log(L)$, yields the fractal dimension in the thermodynamic limit. Note that in the numerical calculation of this section, we consider the averaged fractal dimension (FD) of all eigenstates in pure phases without mobility edges ($V_0 = 0$). In the presence of mobility edges ($V_0 = J$), it corresponds to the averaged FD of

8

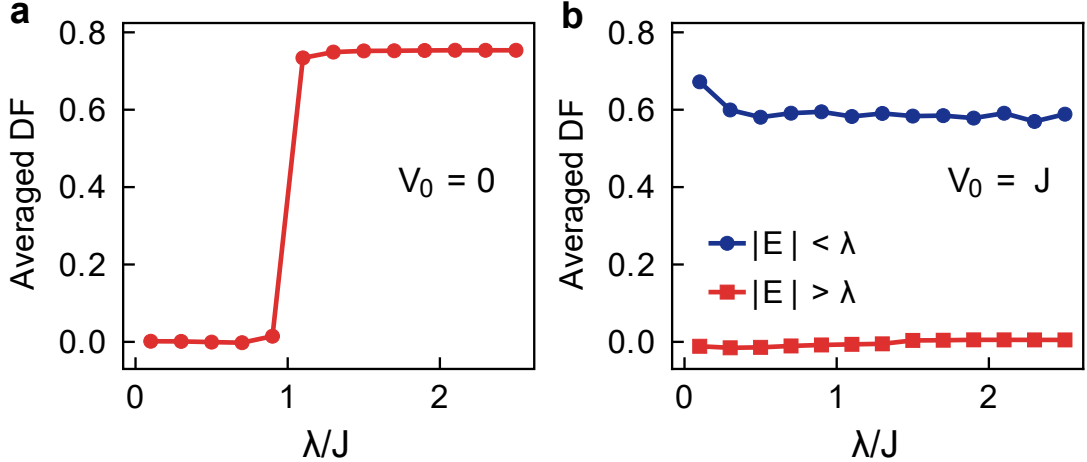

Supplementary Fig. S4. **Fractal dimension at thermodynamic limit for varying $\lambda/J$.** The fractal dimension FD of the eigenstates is determined by a linear fit of the averaged $-\log(\text{IPR})$ versus $\log(L)$. **a** For the mosaic model with $V_0=0$, a phase transition occurs between the localized and critical phases as $\lambda/J$ varies, with FD shows a transition from zero in the localized phase ($\lambda/J < 1$) to approximately 0.74 in the critical phase ($\lambda/J > 1$). **b** For the mosaic model with $V_0 = J$, the presence of mobility edges $E = \pm\lambda$ separates the localized and critical states. In this case, the FD remains nearly zero in the localized regime, while decreases to a value of approximately 0.6 in the critical regime.

all eigenstates within the localized and critical regimes.

We first present the finite size scaling results in Fig. S3 for the case of $\lambda = J/4$ and $\lambda = 2J$ while keeping $V_0 = 0$, corresponding to the localized phase and critical phase, respectively, which are the parameter regimes of Fig. 3a in the main text. To provide a comprehensive analysis, we also include the finite size scaling results for the extended phase, modeled using a Hamiltonian consisting of uniform nearest-neighbor hopping terms. As the system size increases towards the thermodynamic limit, the FD for the localized (extended) phase is fitted to approximately 0 (1), respectively. And the FD for the critical phase is fitted to approximately 0.74.

Next, we consider the FD in the thermodynamic limit for the mosaic lattice model under two conditions: without an on-site quasiperiodic potential ($V_0 = 0$) and with a quasiperiodic potential ($V_0 = J$) for different $\lambda/J$. For the mosaic lattice model without on-site modulation, the system undergoes a phase transition between a localized phase ($\lambda/J < 1$)

9

and a critical phase ($\lambda/J > 1$). Fig. S4a shows that the averaged FD is nearly zero in the localized phase, while it remains consistently around 0.74 in the critical phase for different values of $\lambda/J > 1$, which shows that the FD remains invariant within the critical phase as $\lambda/J$ changes. Similarly, for the mosaic model with mobility edges ($V_0 = J$), the FD in the critical regime remains stable across varying $\lambda/J$, as shown in Fig. S4b. Here, the FD saturates at a lower value of approximately 0.6 when $\lambda/J > 0.1$, compared to the case without a quasiperiodic potential shown in Fig. S4a. This reduction of FD can be attributed to the quasiperiodic on-site potential, which limits the spatial extent of the critical states, leading to more localized peaks in each subregion of the wave functions while preserving their multifractal structure and preventing a transition to a fully localized state.

## II. CHARACTERIZATION OF INCOMMENSURATELY DISTRIBUTED ZE-ORS WITH LONG-RANGE COUPLING

In the main text, we demonstrate experimentally that critical states persist under weak long-range coupling and undergo a transition to extended states only when the coupling strength exceeds a threshold magnitude, which removes all the IDZs. To futher elucidate this phenomenon, we first model large systems using next-nearest neighbor (NNN) hopping terms as a minimal representation of long-range interactions. Numerical simulations confirm that critical states remain robust until the NNN hopping amplitude reaches a threshold value.

We further use the renormalization group (RG) analysis to systematically characterize the transition from critical states to extended state in the presence of long-range coupling. We first show the robustness of the critical states protected by the IDZz, by showing a finite transition NNN hopping coupling strength. This indicates that critical states will be driven into extended states only when NNN coupling exceeds a finite threshold. To explain the smaller transition long-range coupling observed in the main text, we further introduce the next-next-nearest neighbor (NNNN) coupling. The RG calculation shows that, in the presence of the NNNN coupling, the transition threshold of NNN coupling is reduced. This explains the smaller transition long-range coupling in the main text since the 2D superconducting qubit systems involve various ranges of long-range coupling.

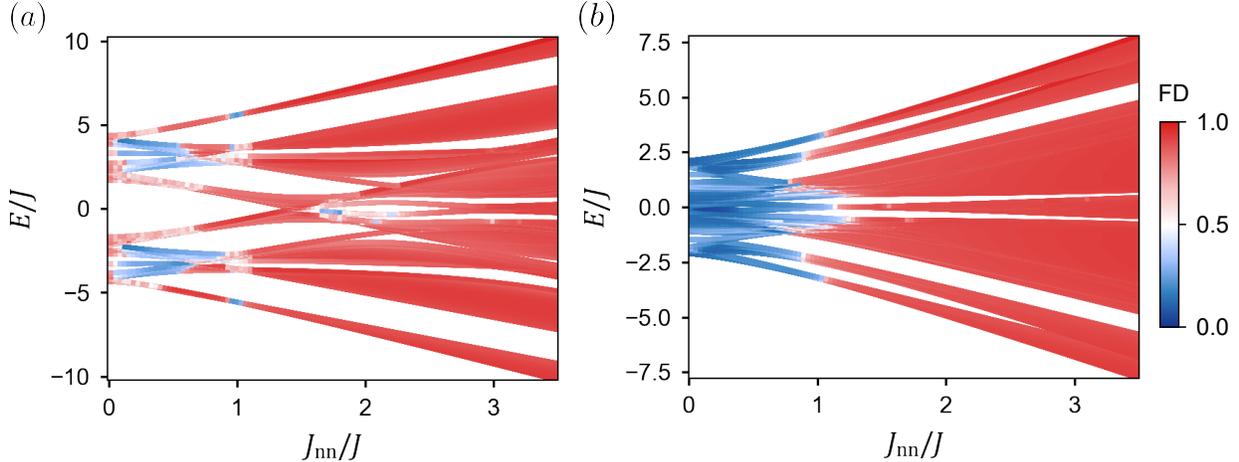

Supplementary Fig. S5. **Mosaic model with long-range coupling. a** Fractal dimension as a function of energy $E/J$ and the next nearest coupling $J_{nn}/J$. Here $\lambda/J = 3$ and system is in critical phase when $J_{nn} = 0$. For such finite but large system size, the critical state will be driven to extended state when $J_{nn}/J$ is large enough. **b** Fractal dimension as a function of energy $E/J$ and the next nearest coupling $J_{nn}/J$. $\lambda/J = 0.5$ and the system is in localized phase when $J_{nn} = 0$. Similarly, system persists in localized phase until $J_{nn}$ is larger than a threshold. For both cases, the system sizes are $L = 2584$.

### A. Numerical demonstration

To show how the IDZs preserve in the presence of long-range coupling, and as a result, give rise to critical states. Here we consider the 1D mosaic lattice model, with the minimal long-range coupling modeled by the next-nearest neighbor (NNN) coupling $J_{nn}$, the Hamiltonian is then given by

$$H = \sum_j J_j(\sigma_j^\dagger \sigma_{j+1}^- + \sigma_{j+1}^\dagger \sigma_j^-) + \sum_j J_{nn}(\sigma_j^\dagger \sigma_{j+2}^- + \sigma_{j+2}^\dagger \sigma_j^-), \tag{S21}$$

with the $J_i$ in a mosaic manner, namely $J_j = \lambda$ for $j$ is odd and $J_j = 2J\cos(2\pi\alpha j + \theta)$ for $j$ is even.

We first start from the critical phase, namely we introduce NNN coupling on top of the critical state by considering the case $\lambda = 3J$. As expected, when $J_{nn}/J$ is not so large, then the system still contains critical states, as indicated in Fig. S5(a). And for the sufficient large $J_{nn}/J$, then the system enters the extended phase.

We then start from the localized phase with $\lambda = 0.5J$, then turn on the $J_{nn}$ also does not

11

drive the localized state into extended state immediately, only when it exceeds a threshold as shown in Fig. S5(b). Moreover, in this case, there no critical states in the whole spectrum. This also manifests the mechanism that, the critical states are generated by the IDZs on top of the delocalized states.

### B. Renormalization group for the long-range mosaic model

#### 1. Review of the renormalization group approach

We first briefly review the procedure of the renormalization group approach. We consider a generic 1D Hamiltonian $H/\hbar = \sum_j t_j(\sigma_j^+\sigma_{j+1}^- + \sigma_j^-\sigma_{j+1}^+) + V_j\sigma_j^+\sigma_j^-$ with quasiperiodic modulation $\cos(2\pi\alpha j + \theta)$, here the irrational number $\alpha$ is an irrational number that can be approached by Diophantine approximation, for example, the golden ratio number that can be approached by the Fibonacci series $\alpha = F_{n-1}/F_n$, with $F_n$ being the Fibonacci series.

This approach aims at determining which localization phases the given state flows to by calculating the relevant parameter of the associated dispersion. Specifically, we first approximate the incommensurate parameter in the quasiperiodic modulation by a commensurate integer $\alpha = \lim_{n\to\infty} F_{n-1}/F_n$, with $F_n$ being the Fibonacci number is the system size $L$. The original Hamiltonian then becomes periodic and exhibits band structure. Then we introduce the quasi-momentum $\kappa_x$ along $x$-direction by the twisted boundary condition or equivalently threading a flux. And we map the phase offset $\theta$ as the quasi-momentum in $y$-direction as $\kappa_y$. Then one can obtain the corresponding Bloch Hamiltonian $\mathcal{H}^{(n)}(\kappa_x, \kappa_y)$. Then we identify which dispersion within the Bloch Hamiltonian becomes relevant and irrelevant as we iterate the size of the unit cell. Specifically, we shall investigate the characteristic polynomial given by $P^{(n)} = \det[H^{(n)}(\kappa_x, \kappa_y) - E]$ for the state with the energy $E$, which can rewrite the $P^{(n)}$ as the form

$$P^{(n)}(E; \kappa_x, \kappa_y) = t_R^{(n)}(E)\cos(\kappa_x + \kappa_x^0) + V_R^{(n)}(E)\cos(\kappa_y + \kappa_y^0)$$
$$+ C_R^{(n)}(E)\cos(\kappa_x + \tilde{\kappa}_x^0)\cos(\kappa_y + \tilde{\kappa}_y^0)$$
$$+ \epsilon_R^{(n)}(E, \varphi, \kappa) + T_R^{(n)}(E), \tag{S22}$$

where $\epsilon_R^{(n)}$ represents higher harmonic terms. The different phases are characterized by investigating how the parameters evolve as $n \to \infty$ by comparing the coefficients of the

12

| $L$ | $t_{1R}^{(L)}(0)$ | $V_{1R}^{(L)}(0)$ | $t_{2R}^{(L)}(0)$ | $V_{2R}^{(L)}(0)$ | $C_R^{(L)}(0)$ |
|---|---|---|---|---|---|
| 3 | 0 | 0 | $2J_{\text{nn}}^6$ | $-2J^6$ | $4J^3\lambda\left(3J_{\text{nn}}^2 + \lambda^2\right)$ |
| 5 | 0 | 0 | $2J_{\text{nn}}^{10}$ | $-2J^{10}$ | $4J^5\lambda\left(5J_{\text{nn}}^4 + 5J_{\text{nn}}^2\lambda^2 + \lambda^4\right)$ |
| 7 | 0 | 0 | $2J_{\text{nn}}^{14}$ | $-2J^{14}$ | $4J^7\lambda\left(7J_{\text{nn}}^6 + 14J_{\text{nn}}^4\lambda^2 + 7J_{\text{nn}}^2\lambda^4 + \lambda^6\right)$ |

Supplementary Table I. Numerical results for the coefficients of the first three unit cells $L = 3$, $L = 5$ and $L = 7$.

corresponding dispersions. For the extended phase, the on-site potential becomes irrelevant $\left|C_R^{(n)}/t_R^{(n)}\right|, \left|V_R^{(n)}/t_R^{(n)}\right| \to 0$, for the localized phase, we have an irrelevant hopping $\left|C_R^{(n)}/V_R^{(n)}\right|, \left|t_R^{(n)}/V_R^{(n)}\right| \to 0$ and for the critical phase, both hopping and on-site potential are relevant $\left|C_R^{(n)}/V_R^{(n)}\right|, \left|C_R^{(n)}/t_R^{(n)}\right| \geq 1$.

### 2. Renormalization group analysis for the next-nearest neighbor coupling

To show how the IDZs preserve in the presence of long-range coupling, and as a result, give rise to critical states. Here we consider the 1D mosaic lattice model, with the minimal long-range coupling modeled by the next-nearest (NN) neighbor coupling $J_{\text{nn}}$, the Hamiltonian is then given by

$$H/\hbar = \sum_j J_j(\sigma_j^+\sigma_{j+1}^- + \sigma_j^-\sigma_{j+1}^+) + \sum_j J_{\text{nn}}(\sigma_j^+\sigma_{j+2}^- + \sigma_j^-\sigma_{j+2}^+), \quad \text{(S23)}$$

with the $J_j$ in a mosaic manner, namely $J_j = \lambda$ for $j$ is odd and $J_j = 2J\cos(2\pi\alpha j + \theta)$ for $j$ is even. To facilitate the discussion, we first consider the transition from critical to extended state of zero energy state, and then generalize it to the finite energy. For $E = 0$, the characteristic polynomial is given by

$$\begin{aligned}P^{(L)}(0;\varphi,\kappa) &= t_{1R}^{(L)}\cos(\kappa_x + \kappa_x^0) + V_{1R}^{(L)}\cos(\kappa_y + \kappa_y^0) \\ &+ t_{2R}^{(L)}\cos[2(\kappa_x + \kappa_x^0)] + V_{2R}^{(L)}\cos[2(\kappa_y + \kappa_y^0)] \\ &+ C_R^{(L)}\cos(\kappa_x + \tilde\kappa_x^0)\cos(\kappa_y + \tilde\kappa_y^0) + \epsilon_R^{(n)}(E=0,\varphi,\kappa) + T_R^{(n)}(E=0).\end{aligned} \quad \text{(S24)}$$

Here $L$ is the size of the commensurately approximated unit cell $L = F_n$. The characteristic

polynomial for any generic approximated unit cell $L$ at for the energy $E$ is given by

$$P^{(L)}(E;\varphi,\kappa) = \det \begin{pmatrix} M_0(E) & \Pi_x^\dagger & 0 & \cdots & \Pi_x \\ \Pi_x & M_1(E) & \ddots & \ddots & \vdots \\ 0 & \ddots & \ddots & \ddots & 0 \\ \vdots & \ddots & \ddots & \ddots & \Pi_x^\dagger \\ \Pi_x^\dagger & \cdots & 0 & \Pi_x & M_{L-1}(E) \end{pmatrix}, \quad (S25)$$

with the $M_j(E)$ and $\Pi_x$ being two-by-two matrices, which are given by

$$M_j(E) = \begin{pmatrix} 2J\cos[2\pi(j-1)\alpha + k_y] - E & 0 \\ 0 & -2J\cos[2\pi(j-1)\alpha + k_y] - E \end{pmatrix}. \quad (S26)$$

and

$$\Pi_x = \begin{pmatrix} e^{-ik_x}(J_{\rm nn} - \frac{\lambda}{2}) & \frac{1}{2}e^{-ik_x}\lambda \\ -\frac{1}{2}e^{-ik_x}\lambda & e^{-ik_x}(J_{\rm nn} + \frac{\lambda}{2}) \end{pmatrix} \quad (S27)$$

Notice that under the transformation $E \to -E$, $k_x \to k_x + \pi$ and $k_y \to k_y + \pi$, the determinant remains the same. Since $L$ is odd, under this transformation, we have $\kappa_x = Lk_x \to \kappa_x + L\pi$ and $\kappa_y = Lk_y \to \kappa_y + L\pi$ altered by an odd number of $\pi$. Therefore, we have

$$\begin{aligned} t_{1R}^{(L)}(-E) &= -t_{1R}^{(L)}(E), \\ V_{1R}^{(L)}(-E) &= -V_{1R}^{(L)}(E), \\ t_{2R}^{(L)}(-E) &= t_{2R}^{(L)}(E), \\ V_{2R}^{(L)}(-E) &= V_{2R}^{(L)}(E), \\ C_R^{(L)}(-E) &= C_R^{(L)}(E). \end{aligned} \quad (S28)$$

Therefore for the zero eigenenergy $E = 0$, we have

$$t_{1R}^{(L)}(0) = 0, \quad V_{1R}^{(L)}(0) = 0. \quad (S29)$$

From the renormalized on-site coupling, we can see to accumulate a $k_y$-dependence of $e^{i2Lk_y}$, the only choice is to multiple $2L$ diagonal terms together, therefore we have the normalized on-site potential strength

$$V_{2R}^{(L)}(0) = -2J^{2L}. \quad (S30)$$

Similarly, for the renormalized hopping coupling, to accumulate a $k_x$-dependence of $e^{i2Lk_x}$, one must multiply $2L$ terms with $e^{ik}$ together. By choosing any $-e^{ik}\lambda$ term, the possibility

14

of choosing two $e^{ik}J_{\text{nn}}$ terms is eliminated, then one cannot find $2L$ terms with $e^{ik}$ to be multiplied. Therefore the only choice is to multiply $2L$ $e^{ik}J_{\text{nn}}$ terms together, which gives normalized hopping coefficient as

$$t_{2R}^{(L)}(0) = 2J_{\text{nn}}^{2L}. \tag{S31}$$

The renormalized hopping coupling for the critical states $C_R^{(L)}(0)$ can be understood via incoherent superposition between different contributions. To obtain such $k_X$ and $k_y$ dependence, one will have $L$ terms with $e^{ik_X}$ and $L$ terms with $e^{ik_y}$ multiplied together. Thus one must multiply $L$ factors of $\pm \frac{J}{2} e^{i(k_y + 2p\pi\alpha)}$ terms and a total of $L$ factors of $e^{ik_x} J_{\text{nn}}$ or $L$ factors of $-e^{ik_x} \lambda$ terms together. Therefore each contribution to the total factor has a random phase factor due to the incommensurate $\alpha$, resulting in an incoherent superposition deducting the combinatorial multiplicity to $\mathcal{O}(1)$ under exponential order at $L \to \infty$. Then since the infinite summation of a geometric progression is of order $\mathcal{O}(1)$, the value at $L \to \infty$ is determined by taking as much only $e^{ik} J_{\text{nn}}$ or only $-e^{ik}\lambda$ terms as possible. Therefore, we obtain the renormalized hopping coupling for the critical states as

$$C_R^{(L)}(0) \sim J^L \left( J_{\text{nn}}^L + \lambda^L \right). \tag{S32}$$

The condition for the emergence of extended state in the presence of next-nearest coupling can be identified by $V_{2R}^{(L)}(0)/t_{2R}^{(L)}(0) \to 0$ and $C(0)/t_{2R}^{(L)}(0) \to 0$, therefore the critical state at $E = 0$ will be driven into the extended state if

$$J_{\text{nn}} > \max\left(J, \sqrt{J\lambda}\right), \tag{S33}$$

which indicates that only if the next-nearest coupling is larger than a certain threshold, the critical states will be driven into the extended states.

### 3. Inclusion of next-next-nearest neighbor coupling

In this subsubsection, we further show that introducing the next-next-nearest neighbor coupling will decrease the transition threshold for $J_{\text{nn}}$ obtained in Eq. (S33). For simplicity of the RG calculation, we introduce the next-next-nearest neighbor coupling on the even site, whose Hamiltonian is given by

$$H = -\mu \sum_{j=\text{even}} (\sigma_{j+3}^+ \sigma_j^- + \sigma_j^+ \sigma_{j+3}^-). \tag{S34}$$

15

If we relabel the odd/even site as the A/B sublattice, then the eigenvalue equation $H|\psi\rangle = E|\psi\rangle$, with $|\psi\rangle = \sum_{j,s=\{A,B\}} u_{j,s}$ takes the form

$$\begin{pmatrix} J_{\text{nn}} & -\lambda \\ -\mu & J_{\text{nn}} \end{pmatrix} u_{j-1} + \begin{pmatrix} J_{\text{nn}} & -\mu \\ -\lambda & J_{\text{nn}} \end{pmatrix} u_{j+1} + \begin{pmatrix} 0 & 2J\cos(2\pi\alpha j) \\ 2J\cos(2\pi\alpha j) & 0 \end{pmatrix} u_j = E u_j, \quad (S35)$$

with a unitary transformation, the eigenvalue equation becomes

$$\begin{pmatrix} J_{\text{nn}} - t_+ & -t_- \\ t_- & J_{\text{nn}} + t_+ \end{pmatrix} u_{j-1} + \begin{pmatrix} J_{\text{nn}} - t_+ & t_- \\ -t_- & J_{\text{nn}} + t_+ \end{pmatrix} u_{j+1} +$$
$$+ \begin{pmatrix} 2J\cos(2\pi\alpha j) & 0 \\ 0 & -2J\cos(2\pi\alpha j) \end{pmatrix} u_j = E u_j, \quad (S36)$$

with $t_\pm = (\lambda \pm \mu)/2$. Then the characteristic polynomial takes the same form as Eq. (S25), with the same $M_j(E)$ as the case $\mu = 0$, and the modified $\Pi_x$ as

$$\Pi_x = \begin{pmatrix} e^{-ik_x\left(J_{\text{nn}} - \frac{\lambda+\mu}{2}\right)} & e^{-ik_x \frac{\lambda-\mu}{2}} \\ -e^{-ik_x \frac{\lambda-\mu}{2}} & e^{-ik_x\left(J_{\text{nn}} + \frac{\lambda+\mu}{2}\right)} \end{pmatrix}. \quad (S37)$$

Following the same logic, for the zero eigenenergy $E = 0$, we have

$$t_{1R}^{(L)}(0) = 0, \quad V_{1R}^{(L)}(0) = 0, \quad V_{2R}^{(L)}(0) = -2J^{2L}. \quad (S38)$$

The renormalized hopping coupling is contributed from the product of either $J_{\text{nn}} \times J_{\text{nn}}$ or the term $(-\lambda)(-\mu)$, which gives

$$t_{2R}^{(L)}(0) = 2\sum_{l=0}^{L} C_L^l J_{\text{nn}}^2(-\lambda)(-\mu) = 2\sqrt{J_{\text{nn}}^2 + \lambda\mu}^{2L} \quad (S39)$$

The renormalized hopping coupling for the critical states $C_R^L$ can be obtained from the same logic in the last subsubsection, which is given by

$$C_R^{(L)}(0) \sim J^L \left(J_{\text{nn}}^L + \lambda^L + \mu^L\right). \quad (S40)$$

From the results Eq. (S38), Eq. (S39) and Eq. (S40), we can see the next-next-nearest neighbor coupling further decreases the transition threshold of $J_{\text{nn}}$ for extended states. In particular, when $\mu$ is small, i.e. $\mu < \lambda$, one finds that $C_R^{(1)}(0)$ under large $L$ limit is independent on $\mu$, yet $t_{2R}^{(2)}(0)$ increases with $\mu$, thus one requires a smaller $J_{\text{nn}}$ to drive

16

critical states into extended states. More precisely, the condition for the emergence of extended state is

$$\frac{V_{t_{2R}}^{(L)}(0)}{t_{2R}^{(L)}(0)} \to 0, \quad \frac{C_{t_{2R}}^{(L)}(0)}{t_{2R}^{(L)}(0)} \to 0 \tag{S41}$$

which is to solve the condition

$$\frac{J^{2L}}{\sqrt{J_{nn}^2 + \lambda\mu}^{2L}} \to 0, \quad \frac{J^L(J_{nn}^L + \lambda^L + \mu^L)}{2\sqrt{J_{nn}^2 + \lambda\mu}^{2L}} \to 0 \tag{S42}$$

at $L \to \infty$, this gives

$$J_{nn} \in \left(\sqrt{J\max(J,\lambda,\mu) - \lambda\mu}, \max(\lambda,\mu)\right) \bigcup \left(\max\left[\sqrt{J^2 - \lambda\mu}, \lambda, \mu\right], \infty\right), \tag{S43}$$

which can be unified as

$$J_{nn} > \sqrt{J\max(J,\lambda,\mu) - \lambda\mu}. \tag{S44}$$

This shows that introducing the next-next-nearest neighbor coupling $\mu$ will modify the transition condition [Eq. (S33)] at $\mu = 0$, indicating the long-range coupling further decrease the transition threshold of $J_{nn}$.

## III. ADDITIONAL INFORMATION FOR THE DYNAMICAL CHARACTERIZATION

This section provides supplementary details regarding the time evolution presented in the main text. We first review the properties of several key quantities used to characterize the dynamics of critical states, and then introduce specific generalizations tailored to our superconducting qubit systems.

In the main text, we analyze the time evolution of the fractal dimension, or equivalently, the dynamical fractal dimension $\mathcal{D}(t)$. To distinguish it from the fractal dimension (FD) obtained from the eigenstates of the Hamiltonian, the dynamical fractal dimension is defined as:

$$\mathcal{D}(t) = -\frac{\log \sum_j |u_{m,j}(t)|^4}{\log N}, \tag{S45}$$

where the $u_{m,j}(t)$ is the time dependent coefficients of the state in the real-space basis, which is given by $|\psi(t)\rangle = \sum_{j=1}^N u_{m,j}\sigma_j^+|\text{vac}\rangle$. The dynamical fractal dimension is closely related to the second-order participation entropy $S_2$ [S5], up to a constant, where:

$$S_2 = -\log \sum_{j=1}^N |u_{m,j}(t)|^4. \tag{S46}$$

17

Both the dynamical fractal dimension $\mathcal{D}(t)$ and dynamical participation entropy $S_2(t)$ quantify the extent to which a quantum state spans the real space. These definitions can be generalized to many-body systems to measure the spread of quantum states in the Hilbert space [S6].

To probe the phase diagram of the system described in the main text, we utilize the time-averaged dynamical fractal dimension, defined as:

$$\overline{\mathcal{D}} = \frac{1}{t_f} \int_0^{t_f} [\mathcal{D}(\tau) - \mathcal{D}(0)] d\tau, \tag{S47}$$

which smooths out oscillations over time and serves as a reliable indicator of the phase of the system.

### A. Wave packet dynamics

Wave packet dynamics is a widely used method to characterize the dynamical behavior of quantum states. The localized, extended, and critical phases can be distinguished by monitoring the time evolution of an initial wave packet, typically initialized as a Gaussian distribution with half-width $a$, centered at the site $j_0$:

$$\psi_j(t=0) = \frac{1}{\sqrt{\sqrt{\pi}a}} e^{-(j-j_0)^2/2a^2}, \tag{S48}$$

The evolution can be characterized by the mean square displacement $W(t)$, which measures the width of the wave packet and is defined as:

$$W(t) = \left[ \sum_j (j-j_0)^2 |\langle \psi(t) | j \rangle|^2 \right]^{1/2}. \tag{S49}$$

In the long time limit, $W(t)$ exhibits universal scaling:

$$W(t) \sim t^\kappa, \tag{S50}$$

where the dynamical exponent $\kappa$ takes characteristic values depending on the phase: $\kappa = 1$ for extended phase, $k = 0$ for localized phase, and $0 < \kappa < 1$ for critical phase [S7–S9].

In our superconducting qubit implementation, we replace the initial Gaussian wave packet with a single-site occupation. To improve robustness against local noise in experiments, we redefine the mean square width using:

$$W(\tau) = \sum_j \sqrt{|j-j_0|} |\langle \psi(\tau) | j \rangle|^2. \tag{S51}$$

Instead of focusing on the long-time universal behavior of $W(\tau)$, we analyze its early-time dynamics as a complementary marker for the dynamical fractal dimension. Specifically, we calculate the integrated width:

$$M(t_f) = \frac{1}{t_f} \int_0^{t_f} [W(\tau) - W(0)] d\tau, \tag{S52}$$

which captures the early-time spin dynamics in the system.

## IV. DEVICE FABRICATION AND ASSEMBLY

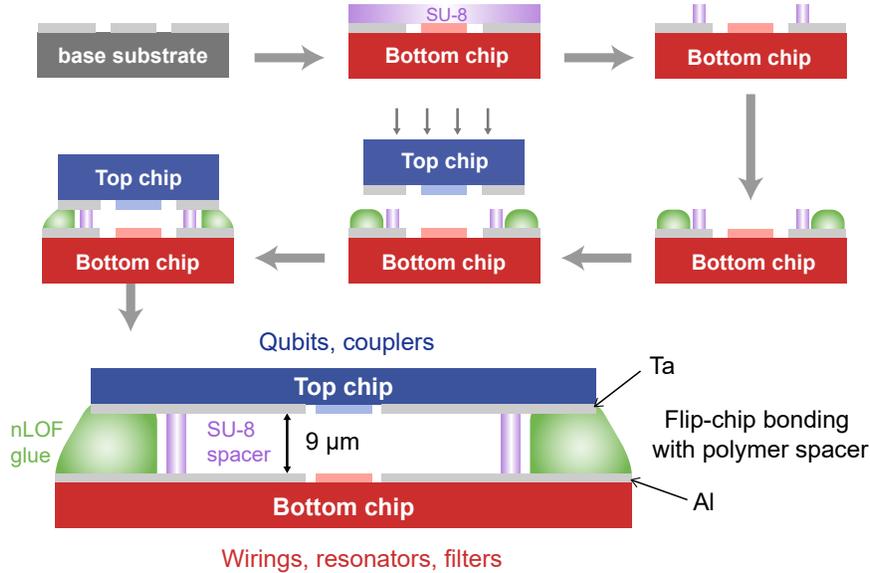

Supplementary Fig. S6. **Fabrication process of the quantum processor.** The processor consists of a top qubit chip and a bottom carrier chip, bonded face-to-face using SU-8 and nLOF glue.

The experiment in this work is carried out on a two-dimensional (2D) superconducting quantum processor consisting of 66 frequency-tunable transmon qubits and 110 tunable couplers. The processor comprises a top chip and a bottom carrier chip, bonded face-to-face using SU-8 and nLOF glue [S10]. The top chip hosts the qubits and couplers, whereas the bottom carrier chip hosts the readout resonators as well as control and readout wiring circuitries. The fabrication and assembly process of the quantum processor, as illustrated in Fig. S6, involves the following steps:

1. A 100 nm aluminum is deposited onto a sapphire wafer using electron beam evaporation for the bottom chip; and a 100 nm tantalum film is deposited onto another sapphire wafer using sputtering for the top chip respectively.

2. Large-scale structures, including the control and readout circuits on the bottom chip, as well as the capacitor pads for the qubits and couplers on the top chip, are realized through optical lithography and subsequent wet etching.

3. To mitigate signal crosstalk, $SiO_2$-supported bridges are created on the bottom chip to shield critical circuits.

4. The Al-$AlO_x$-Al Josephson junctions are patterned on the top chip via electron beam lithography and fabricated using the double-angle electron beam evaporation.

5. Bandage technology [S11] is employed to establish a galvanic connection between the aluminum junctions deposited in step 4 and the tantalum film deposited in step 1.

6. 9-$\mu$m-tall SU-8 photoresist is positioned at the corners of the bottom carrier chip as a spacer between the top and bottom chips, and then the top and bottom chips are bonded together using nLOF glue.

The device fabricated using the above technique is robust in performance after several cycles of cooling down and warming up, and the bonding process of two separate chips through polymer spacers enables us to recycle the bottom (top) chip if the performance of the other chip is not good after cooling down and being measured, which increases the productivity of our fabrication.

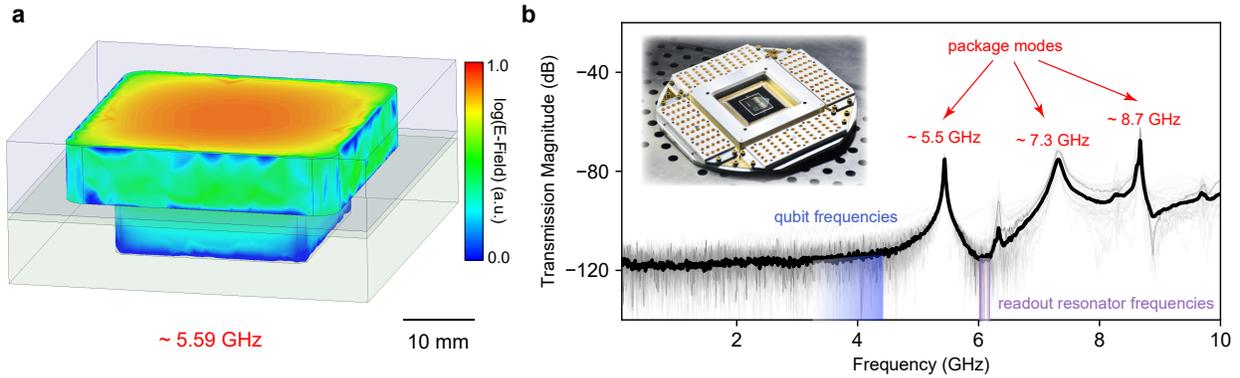

Supplementary Fig. S7. **Package modes. a** Example EM simulation of the box model revealing the package mode at 5.59 GHz (others larger than 7.0 GHz). The box modes in real device can slightly vary as the bonding connections are involved. **b** The transmission magnitudes for the device package are measured by vector network analyzer via multiple ports (grey lines) and being averaged (black line). The probed package modes are marked by the red arrows, which are detuned from the working frequencies of qubits and readout resonator. The inset (upper left) shows an example photograph of the package with the cover removed.

Care must be taken when designing the package for quantum chips of this scale [S12, S13]. Fig. S7 shows the electromagnetic (EM) simulation and measurement of the package mode for our device. EM simulation of the package, as shown in Fig. S7a, reveals the lowest package mode at 5.59 GHz. In Fig. S7b, we measure the transmission magnitudes for the device package by vector network analyzer via multiple ports. The lowest package mode is probed at 5.5 GHz, which is close to the simulated value. The slight deviation could be due to the bonding connections involved in the real device. The working frequencies of the qubits ($\sim 4.2$ GHz) are well below the fundamental box mode. The readout resonators ($\sim 6.2$ GHz) are strategically positioned between the fundamental box mode at 5.5 GHz and the secondary mode at 7.3 GHz.

## V. EXPERIMENTAL SETUP

The quantum processor is mounted on the mixing chamber plate of a dilution refrigerator (DR) with a temperature of around 10 mK, as shown in Fig. S8a. The room-temperature and cryogenic wirings in our experimental setup are illustrated in Fig. S8b, in which we use custom-made digital-to-analog converter (DAC) and analog-to-digital converter (ADC) circuit boards for qubit control and measurement, respectively. In the quantum processor, the 66 qubits are arranged in a square lattice, from which we select 56 qubits arranged in a 1D array, as shown in Fig. S11a. Each nearest-neighbor (NN) qubit pair is connected by a tunable coupler to control the effective coupling strength between qubits. The qubits have two asymmetric Josephson junctions with $E_{J1}/E_{J2} = 3.6$, where $E_{J1}$ and $E_{J2}$ are the Josephson energies of the two junctions. The frequency of each qubit can be individually adjusted by varying the corresponding external flux through the Z control line and ranges from approximately 3.9 GHz to 5.2 GHz, as shown in Fig. S9 and Fig. S11b. Typical qubit relaxation time $T_1$ at their idle frequencies are shown in Fig. S11c. The state of the qubit can be deduced by measuring the state-dependent transmission of the readout resonator using the dispersive readout scheme, where the dedicated readout resonator with frequency around 6.15 GHz is coupled to each qubit. Fig. S10 and Fig. S11d display the qubit readout fidelities, with median values of 0.96 for the $|0\rangle$ state, 0.93 for the $|1\rangle$ state, respectively.

The two floating tunable transmon qubits in our device are capacitively coupled to a floating tunable coupler [S14], and each floating qubit is surrounded by four couplers [S15].

22

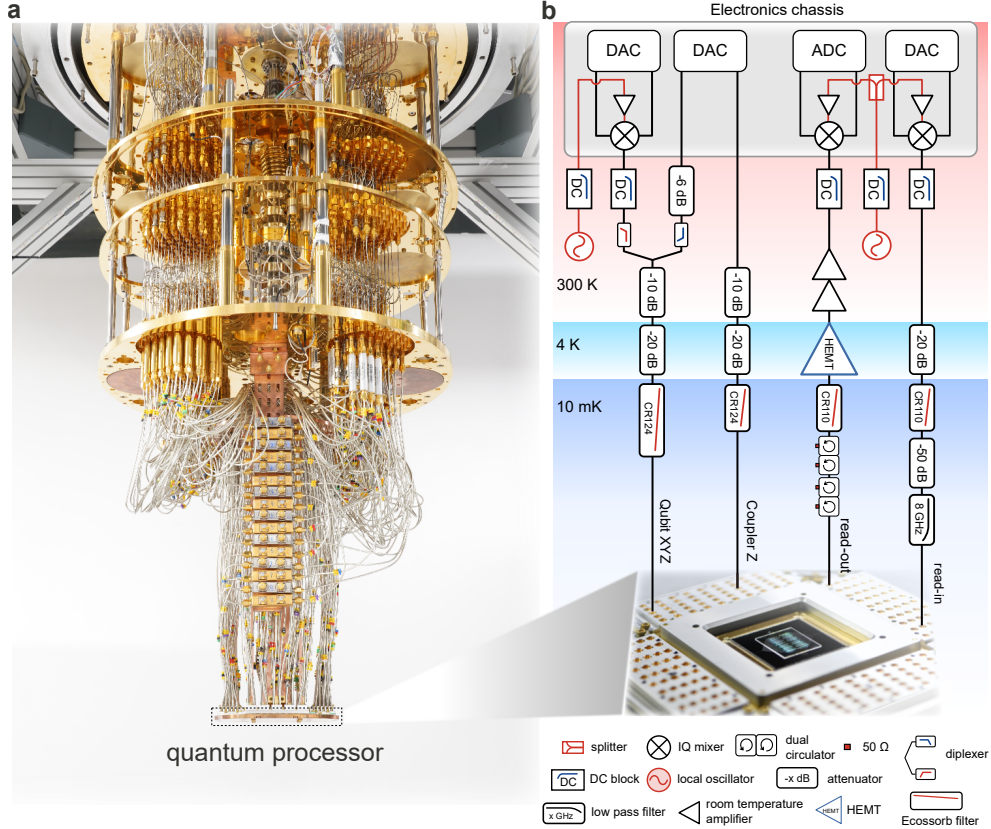

Supplementary Fig. S8. **Experimental setup. a** Photograph of the cryogenic setup with the mounted superconducting quantum processor. **b** Illustration for room-temperature and cryogenic wiring.

The circuit schematic of the qubit-coupler-qubit system is shown in Fig. S12a, where the effective coupling strength between qubits $Q_A$ and $Q_B$ can be controlled by applying external flux on the corresponding coupler $C$ [S16, S17]. The system of qubits $Q_{A,B}$ and coupler $C$ can be described by the Hamiltonian

$$H/\hbar = \sum_{i=A,B,C} \left( \omega_i a_i^\dagger a_i + \frac{U_i}{2} a_i^\dagger a_i^\dagger a_i a_i \right) + g_{AC} \left( a_A^\dagger a_C + a_A a_C^\dagger \right) + g_{BC} \left( a_B^\dagger a_C + a_B a_C^\dagger \right) \\ + g_{AB} \left( a_A^\dagger a_B + a_A a_B^\dagger \right), \tag{S53}$$

where $a_i$ ($a_i^\dagger$) is the annihilation (creation) operator, $\omega_i$ is the qubit or coupler frequencies, $U_i$ is the qubit or coupler anharmonicities, $g_{AC}$ ($g_{BC}$) is the coupling strength between $Q_A$ ($Q_B$) and coupler $C$, and $g_{AB}$ is the coupling strength between $Q_A$ and $Q_B$. When the coupler frequency is largely detuned from the qubits frequencies $|\omega_{A,B} - \omega_C| \gg g_{AC}, g_{BC}$,

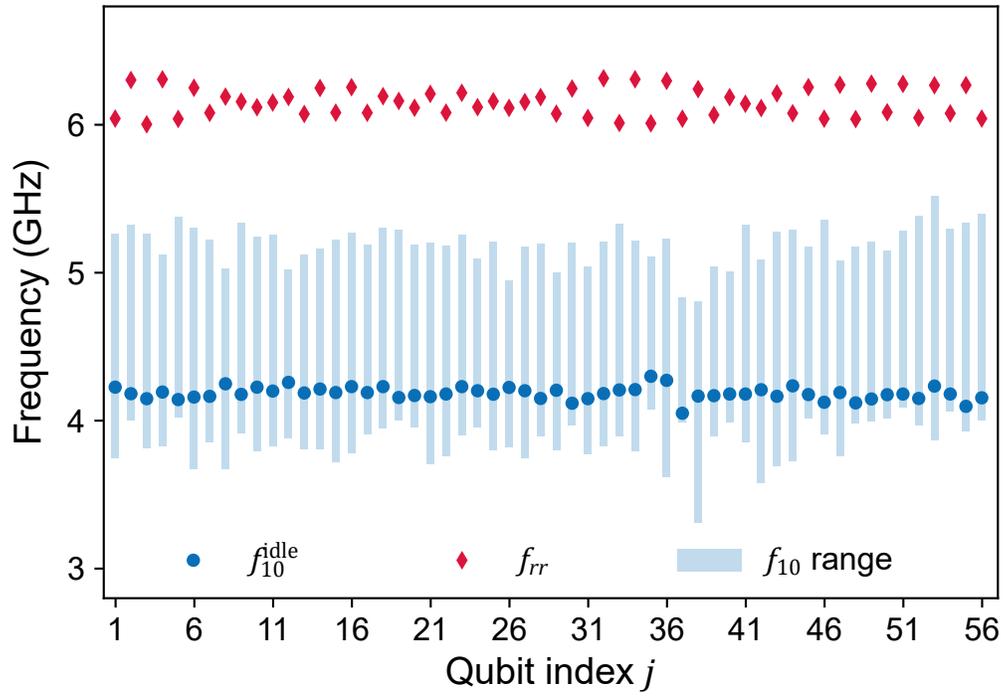

Supplementary Fig. S9. **Characterization of qubit frequencies $f_{10}$ and readout resonator frequencies $f_{rr}$.**

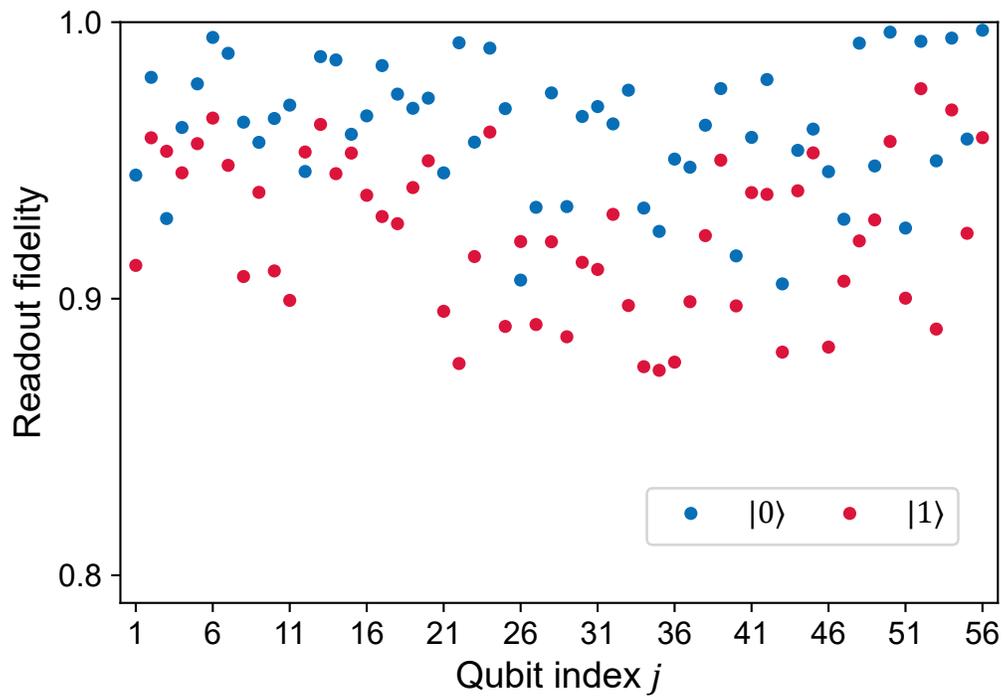

Supplementary Fig. S10. **Characterization of readout fidelities for states $|0\rangle, |1\rangle$.**

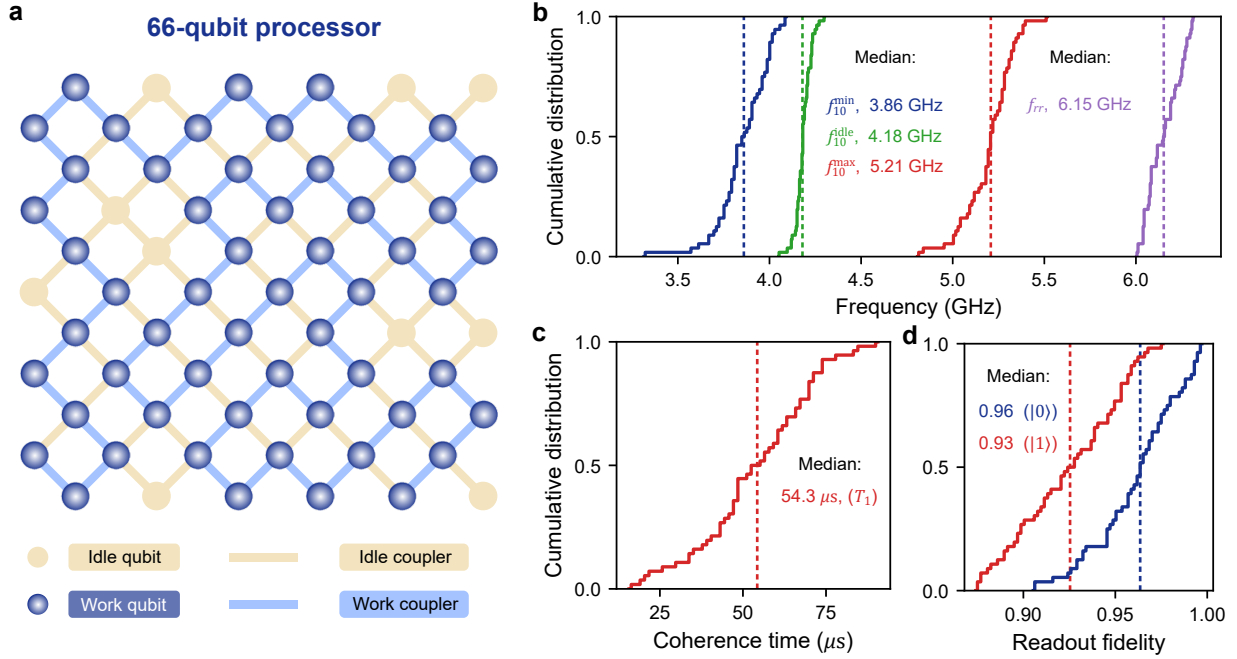

Supplementary Fig. S11. **Device performance.** **a** The layouts of qubits and couplers. **b** Cumulative distribution of the minimum frequency $f_{10}^{\min}$, maximum frequency $f_{10}^{\max}$, and idle frequency $f_{10}^{\text{idle}}$ for the frequency-tunable qubit, and the frequency of readout resonator. **c** Cumulative distribution of qubit relaxation time $T_1$ measured at the idle frequency across the 56 qubits used in experiment. **d** Cumulative distribution of the qubit readout fidelities for $|0\rangle$ and $|1\rangle$ state.

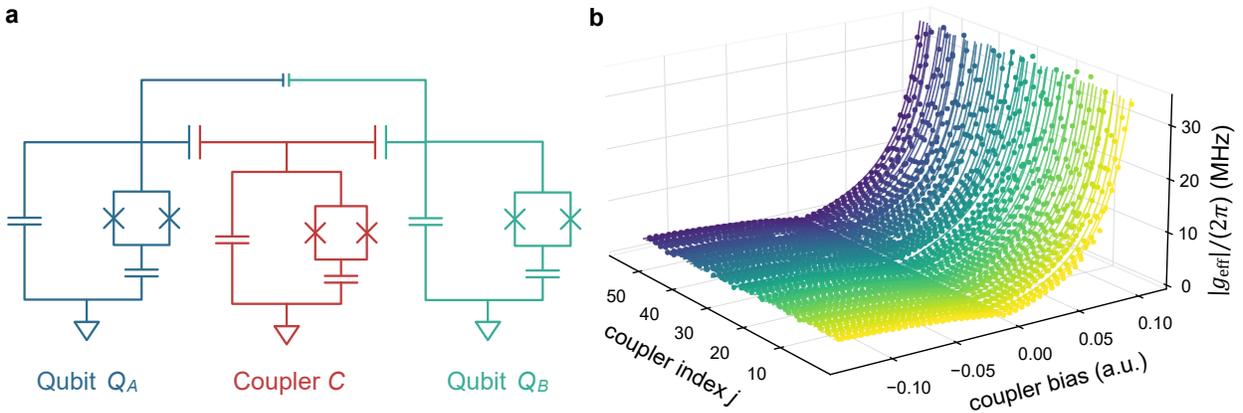

Supplementary Fig. S12. **Characterization of the tunable coupling.** **a** Circuit schematic for the qubit-coupler-qubit system. **b** The coupling strength $|g_{\text{eff}}|$ extracted from the vacuum Rabi oscillation between the neighboring qubits at different coupler bias.

the effective Hamiltonian of such a system is given as

$$H_{\text{eff}}/\hbar = \sum_{i=A,B}\left(\omega_i a_i^\dagger a_i + \frac{U_i}{2}a_i^\dagger a_i^\dagger a_i a_i\right) + g_{\text{eff}}\left(a_A^\dagger a_B + a_A a_B^\dagger\right), \tag{S54}$$

where the effective coupling strength

$$g_{\text{eff}} = g_{AB} + \frac{g_{AC}g_{BC}}{2}\left(\frac{1}{\omega_A - \omega_C} + \frac{1}{\omega_B - \omega_C}\right), \tag{S55}$$

can be modulated by tuning the coupler frequency $\omega_C$ through the flux bias line of the coupler. In experiments, we perform the vacuum Rabi oscillation between the first excited states of two qubits to characterize the effective coupling strength $g_{\text{eff}}$ between qubits, where a Z control pulse is applied on the coupler with different coupler bias to change the strength of $g_{\text{eff}}$. Fig. S12b shows the characterization of the couplers $C_{j,j+1}$ connecting the qubits $Q_j$ and $Q_{j+1}$ in the 1D array, where the effective coupling strength $g_{\text{eff}}$ can be continuously adjusted from $+4$ MHz to approximately $-30$ MHz [S16].

## VI. ADDITIONAL EXPERIMENTAL RESULTS OF THE TIME EVOLUTION

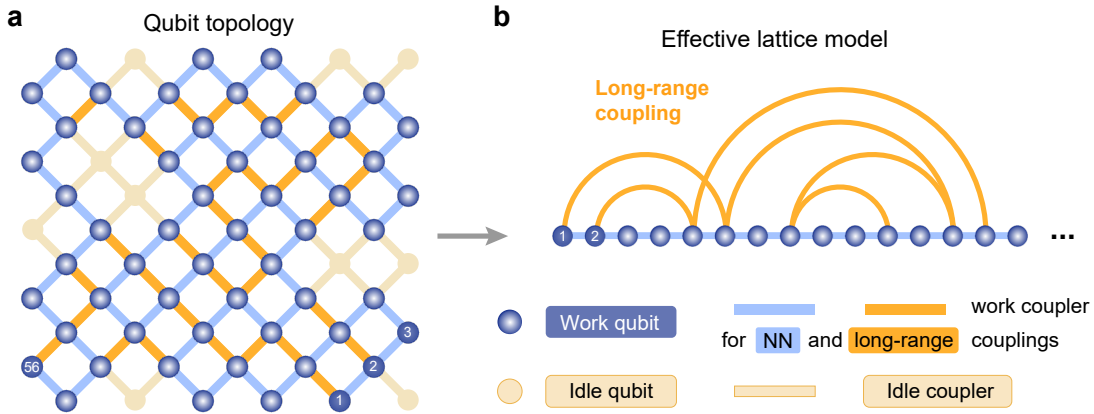

Supplementary Fig. S13. **1D lattice model with long-range coupling on a 2D superconducting quantum processor.** **a** The qubit topology for a 2D superconducting quantum processor. **b** The effective 1D lattice model with long-range coupling.

The 2D superconducting quantum processor enables us to emulate the processes with long-range coupling beyond the original 1D array with various controlled configurations, as shown in Fig. S13, which is used in Fig. 2e and f of the main text. In Fig. 2 of the main text,

the dynamics of the critical state in the mosaic lattice model shows uni-propagating regimes around the site $j = 14$ and $j = 47$. We further investigate the critical state dynamics in the presence of long-range couplings under different configurations of long-range couplings.

We first show that, the IDZs protected critical states is robust to the local perturbative long-range couplings, as long as the quasiperiodic modulation of hopping couplings and the

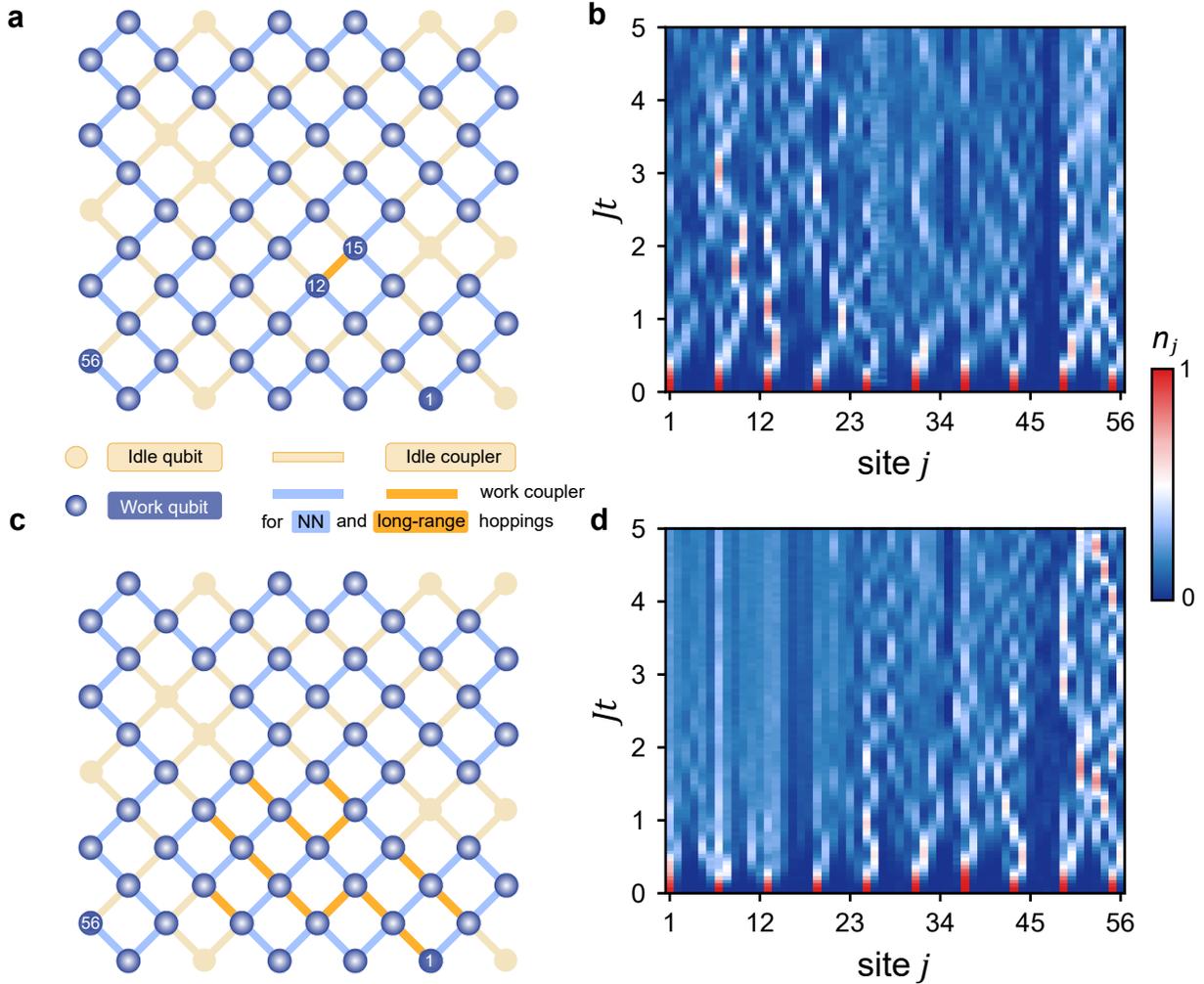

Supplementary Fig. S14. **Dynamics of the mosaic model in the presence of long-range coupling.** **a** Illustration for a 1D array with a single long-range coupling between the sites $12 \leftrightarrow 15$. **b** Measured dynamics in the configuration of a 1D array with long-range couplings as illustrated in **a**. **c** Illustration for a 1D array with more long-range couplings. **d** Measured dynamics in the configuration as illustrated in **b**. Here $\lambda/(2\pi) = 10$ MHz, $J/(2\pi) = 4$ MHz, $J_{m,n}^L/(2\pi) = 10$ MHz and $\phi = \pi/5$.

27

overall IDZs persist. As illustrated in Fig. S14a, when a single long-range coupling between the sites 12 ↔ 15 is involved, only the largest zeros in the quasiperiodic hopping couplings is removed under this local perturbation. Fig. S14b shows the measured dynamics, and the density evolution pattern of critical states still persists. This indicates that, the critical states are robust to the local perturbation of the IDZs.

Introducing more long-range couplings generically breaks the critical states. In Fig. S14c, we introduce the long-range coupling into half of the system, while keeps the IDZs in the rest of the system unchanged. Fig. S14d shows the characteristic dynamics of extended state in the presence of long-range coupling. The unperturbed part, however, the dynamics remains non-ergodic.

---


\end{document}